\documentclass[11pt]{article}
\newcommand{\mcolor}{black}        %Color name for changes made after review
\usepackage{graphicx}
\usepackage{fancyhdr}
\usepackage{makeidx}
\usepackage{amssymb,amsmath}
\usepackage{physics}
\usepackage{upquote}
\usepackage{xcolor}
\usepackage[utf8]{inputenc}
\usepackage[nottoc]{tocbibind}
\usepackage{setspace}
\usepackage{enumerate}
\usepackage{textcomp}
\usepackage{gensymb}
\usepackage{xr}
\usepackage{comment}
\usepackage{lineno}
\usepackage{mathtools}
\usepackage[normalem]{ulem}
\usepackage[export]{adjustbox}
\usepackage{titling}

\usepackage{tcolorbox}
\newcommand{\revised}[1]{\textcolor{black}{#1}}
\usepackage[hidelinks,colorlinks,allcolors=.,urlcolor=black,citecolor=black]{hyperref}
\usepackage[sorting=none,doi=false,url=false,citestyle=numeric-comp,date=year]{biblatex}
\AtEveryBibitem
{
  \clearfield{note}
  \clearfield{issn}
  \clearfield{eprint}
}
\renewbibmacro{in:}{}
\DeclareFieldFormat{pages}{\mkfirstpage[{\mkpageprefix[bookpagination]}]{#1}}
\newbibmacro{string+doi}[1]{%
  \iffieldundef{doi}{#1}{\href{http://dx.doi.org/\thefield{doi}}{#1}}}
\DeclareFieldFormat{title}{\usebibmacro{string+doi}{\mkbibemph{#1}}}
\DeclareFieldFormat[article]{title}{\usebibmacro{string+doi}{\mkbibquote{#1}}}
\newcommand{\vs}{\vspace}
\newcommand{\hs}{\hspace}
\newcommand{\alpham}{\ensuremath{\alpha}}
\newcommand{\betam}{\ensuremath{\beta}}
\newcommand{\taum}{\ensuremath{\tau}}
\newcommand{\degreem}{\ensuremath{\degree}}
\newcommand{\gammam}{\ensuremath{\gamma}}

\newcommand{\Delm}{\ensuremath{\Delta}}

\newcommand{\sigmam}{\ensuremath{\sigma}}

\newcommand{\neqm}{\ensuremath{\neq}}
\newcommand{\pmm}{\ensuremath{\pm}}
\newcommand{\simm}{\ensuremath{\sim}}

\newcommand{\hslashm}{\ensuremath{\hslash}}
\newcommand{\sqrtm}[1]{\ensuremath{\sqrt{#1}}}

\newcommand{\ketm}[1]{\ensuremath{\ket{#1}}}
\newcommand\braketma[2]{\ensuremath{\braket{#1}{#2}}}
\newcommand\braketmb[3]{\ensuremath{\langle{#1}\lvert{#2}\lvert{#3}\rangle}}
\ExplSyntaxOn
\newcommand\braketm[3]
  { \tl_if_empty:nTF {#2} { \tl_if_empty:nTF {#3}{\braketma{#1}{#1}}{\braketma{#1}{#3}} } { \braketmb{#1}{#2}{#3} } }
\ExplSyntaxOff

\newcommand{\fracm}[2]{\ensuremath{\frac{#1}{#2}}}
\newcommand{\gttau}{g\ensuremath{^{(2)}(\tau)}}
\newcommand{\gtzero}{g\ensuremath{^{(2)}(0)}}
\newcommand{\tsup}{\textsuperscript}
\newcommand{\tsub}{\textsubscript}

\usepackage{longtable}  
\usepackage{tabularx}   
\usepackage{booktabs}  
\usepackage{caption}
\usepackage{rotating}
\usepackage{multirow}
\newcommand{\etal}{et al.}

\title{The origin and promise of transition metal dichalcogenide hosted single photon emitters for quantum technologies}
\author{Mayank Chhaperwal, Amartyaraj Kumar, and Kausik Majumdar$^{*}$\\
	$^1$Department of Electrical Communication Engineering, \\Indian Institute of Science, Bangalore 560012, India\\
	$^*$Corresponding author, email: kausikm@iisc.ac.in}
\date{}
\begin{document}

\maketitle

\begin{abstract}
Single photon emitters (SPEs) are integral parts of several quantum technology implementations. Over the past decade or so, monolayers of transition metal dichalcogenides (TMDCs) have emerged as one of the promising candidates for SPE platforms with attractive characteristics. To move ahead, it is necessary to understand the atomistic origin of SPEs in TMDCs - a topic which is highly debated with contradicting proposals. In this paper, we critically review these existing proposals to elucidate their origin. Further, we perform a critical trend analysis for different figures of merit of TMDC-based SPEs, and propose a characterization methodology to streamline the reporting process. Finally, we review several quantum technology implementations where solid state SPEs are being used, and identify the advancements required in TMDC-based SPEs for their successful adoption in these technologies.
\end{abstract}

% \begin{tcolorbox}[ boxrule=0.5pt, colback=white, colframe=black, width=\textwidth]
% \texttt{Disclaimer} \\ \\
% use \verb|\revised{text...}| to add any new text and use \verb|\sout{text...}| to strike through any previous text.  \revised{Example Text is revised here.} \sout{The text is removed.}
% \begin{itemize}
%     \item Commentor 1: No comments
%     \item Commentor 2: a lot of comments. Added in inline comment boxes for better guidance.
% \end{itemize}
% \end{tcolorbox}

% \scriptsize \listoftodos

\normalsize
\newpage
\section*{Introduction}
Single photon emitters (SPEs) are a special case of quantum emitters where emission contains photons in the Fock state with exactly one photon per wavepacket ($\ket{1}$). These are different from other classical sources of light, such as lasers and thermal sources, in their photon emission statistics. While photons emitted from a laser and a thermal source follow Poissonian and super-Poissonian statistics, respectively, the photons emitted from an SPE follow sub-Poissonian statistics. For sub-Poissonian sources, $\Delm{}n < \bar{n}$, where $\bar{n}$ is the average number of photons in a wavepacket and $\Delm{}n$ is the variance. For SPEs, $\bar{n}=1$ and $\Delm{}n = 0$.

SPEs occupy a central position in multiple quantum applications. While early quantum technology predominantly relied on non-linear heralded photon sources, scalability requirements have motivated developments in SPEs. Nowadays, these are the promising contenders for linear-optics quantum computing \revised{(LOQC)} \cite{OBrien2009, Knill2001, maring_versatile_2024}, quantum simulation and quantum walks \cite{wyborski_toward_2025,aspuru-guzik_photonic_2012}, boson sampling \cite{loredo_boson_2017,He2017, Wang2018d, Wang2019}, quantum key distribution \revised{(QKD)} protocols such as BB84 and E91 \cite{bennett_quantum_2014, ekert_quantum_1991, morrison_single-emitter_2023, barnes_decoy-state_2025, zhaoSitecontrolled2021}, quantum teleportation using photonic entanglement \cite{laneve_quantum_2025, strobel_telecom-wavelength_2025, anderson_quantum_2020}, and quantum random number generation \cite{chen_single_2019, hoese_single_2022}. In each of the applications, single photons serve as the fundamental carriers of quantum information. The performance of the platform is thus governed by the quality of the SPE in terms of photon purity, indistinguishability, fidelity, and emitter brightness. Although the relative weight of these metrics varies for different applications, SPEs are showing promise of achieving the desired control over every parameter and the required accuracy for almost every well-protocolized application. \\

Over the last decade, ultra-thin semiconducting layers of transition metal dichalcogenides (TMDCs) have emerged as an intriguing platform for SPEs. Recent developments suggest that the quality of the TMDC-based SPEs is approaching towards the best quality solid state SPEs around, which makes them promising for quantum technology applications. \textcolor{\mcolor}{However, given a relatively nascent field, several scientific debates remain unresolved, including the atomic origin of the SPEs, the role of strain in funnelling excitons and hybridization, the origin of fine-structure splitting of the sharp emission peak, the origin of the observed g-factors, and the charge state of the emitter. Moreover, there is a lack of consensus regarding the precise ways of reporting the results, and also a lack of clarity regarding how exactly the TMDC-based SPE platform can become a viable contender for quantum technology. To this end, we aim to elucidate these aspects in this review.} Accordingly, the rest of the paper is organized as follows:

\revised{At first,} the important figures of merit (FOMs) for SPE are defined. Next, we outline the benefits of TMDCs as a platform for the development of SPEs. The origin of the SPEs in TMDCs remains one of the highly debated topics. To throw some light on this, in \revised{the following section}, we provide a detailed survey of the literature focusing on the origin of SPEs in WSe\tsub{2}. \revised{After that}, we analyze and identify trends in the reported SPE FOMs across multiple implementations in TMDCs. Next, to streamline the reporting process, we propose a characterization procedure for measuring and reporting the crucial FOMs, which will help the benchmarking process in the future. \revised{This is followed by a discussion on} various quantum technologies where SPEs, in general, are being employed. In this section, we also discuss the need for specific improvements in the TMDC-based SPEs to enable their efficient adoption in each of these technologies. Finally, we conclude the paper with an outlook on the future.
\section*{Desirable characteristics of SPE}{\label{SPE_FOM}}

\subsection*{Brightness} Brightness ($B$) of an SPE typically refers to the maximum rate at which it can emit photons. This quantity is correlated with the radiative lifetime of the emitters and can be expressed either as the number of emitted photons per unit time or as the quantum efficiency of the emitter. These two variations are discussed \revised{in the later section on \nameref{Brightness}}.
\subsection*{Purity} 
\begin{figure}[!hbt]
	\centering
	\vs{-0.3in}
	\hs{-0in}
	\includegraphics[width=1.3\paperwidth,center]{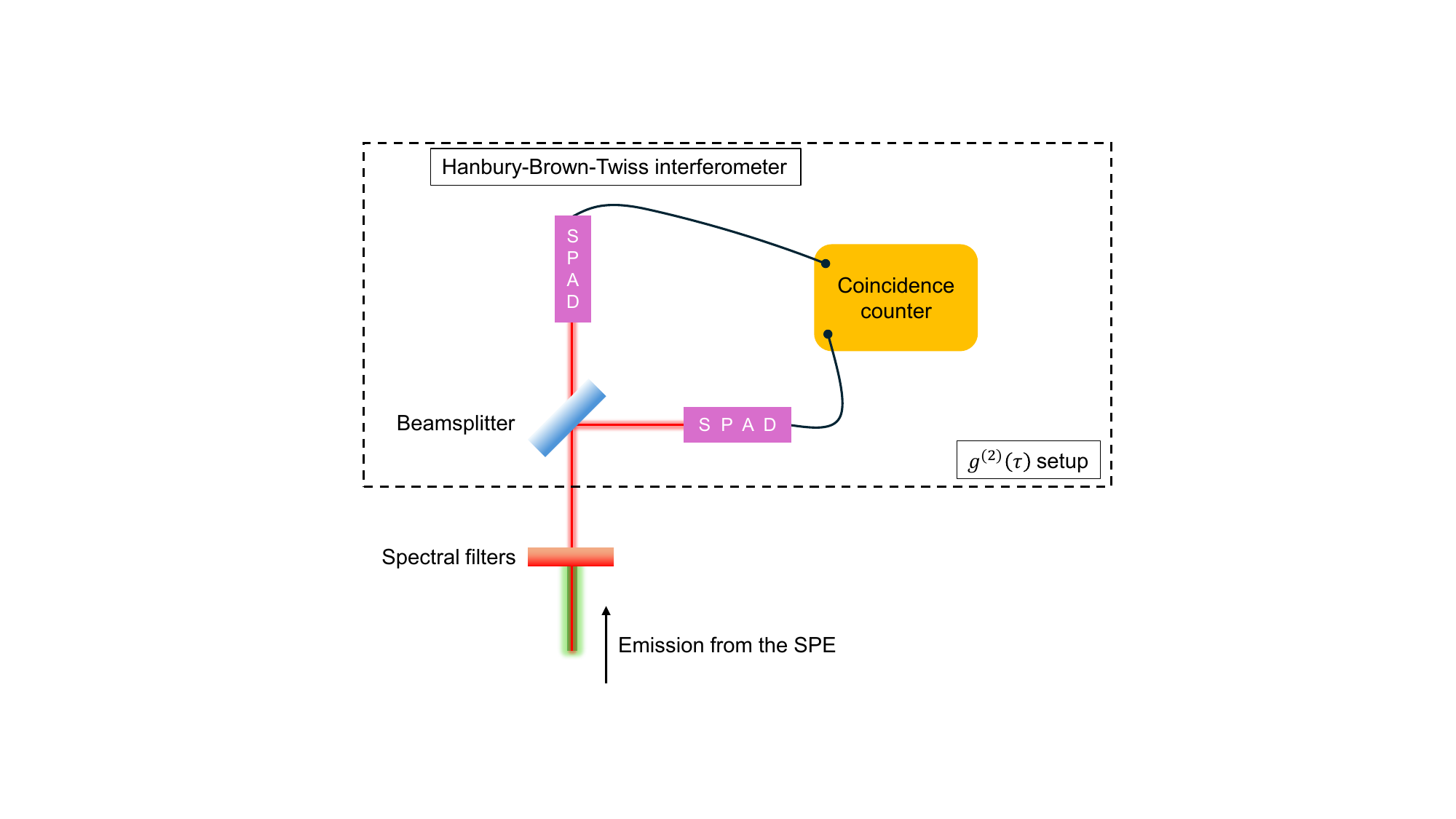}
	\vspace{-1.2in}
	  \caption{\textbf{Schematic of the setup required to measure the second-order correlation function of the SPE.} The black arrow shows the direction of the signal from the sample, which is spectrally filtered to pass only the SPE emission peak. The filtered signal then enters the dashed box, which is the HBT setup containing a 50:50 beam splitter and two SPADs at the two output arms of the beam splitter for the measurement of \gttau{}. Outputs of the SPADs are connected to a coincidence counting module.}
      \label{fig:g2_setup}
\end{figure}

An ideal SPE should have no multiphoton component in its emission. Practically, the true single photon emission is mixed with an uncorrelated background emission, degrading the purity ($P$) of the SPE. This is quantified as $P=1-\gtzero{}$ and found by measuring the second-order correlation function \gttau{} of the emitter in a Hanbury-Brown \& Twiss (HBT) setup (\autoref{fig:g2_setup}). Ideally, a single photon, when impinged on a beam splitter , can be detected at only one of the single photon avalanche detectors (SPADs), and thus the value of coincidence counts for null relative path difference - \gtzero{} - is zero. Specific details about measuring and reporting the purity of the SPEs are discussed in detail \revised{in the later section on \nameref{Purity}}.
\subsection*{Indistinguishability} An important mechanism in many of the photonic implementations of quantum technologies is the interference of single photons. Consecutive photons emitted from the SPE can be made to interfere with each other only if they are identical. Indistinguishability of the SPE quantifies this identical nature. For a source to achieve a high degree of indistinguishability, the photons emitted by it should have a high degree of polarization, the lifetime of the source should be as low as possible, the linewidth of the source should be low, the coherence time of the source should be high, and the source should be free from any spectral/intensity jitter. More details about measuring and reporting the purity of the SPEs are discussed \revised{in the subsequent section on \nameref{Indistinguishability}}.
\subsection*{Temporal determinism} An ideal SPE should be capable of producing a single photon on demand based on an external trigger. This trigger can be an excitation laser pulse or an electrical signal in the case of electrically excited SPE. A perfect quantum efficiency of the emitter is required to achieve fully deterministic SPE, as discussed \revised{in the subsequent section on \nameref{Brightness}}.
\subsection*{Spatial determinism} For rapid employment and scaling of SPEs in quantum technologies, these should ideally be deterministically fabricated at desired locations. This is critical when introducing an SPE in a photonic cavity for improving its performance since the Purcell factor of the cavities strongly depends on the location of the emitter within them.
\subsection*{Collection efficiency} This quantity refers to the ease of outcoupling of the emitted photons with minimum loss in brightness. This is usually reported as the percentage of photons successfully collected into an objective/waveguide/fiber to the total emitted photons by the SPE. A perfect temporal deterministic source behaves as a probabilistic source when the collection efficiency is poor, as discussed \revised{in the subsequent section on \nameref{Brightness}}.
\subsection*{Operation temperature} This refers to the temperature at which the SPE can sustain a high brightness while maintaining high single-photon purity. For practical applications, a higher upper limit on the operating temperature is desirable.
\subsection*{Spectral/Intensity jitter} Due to external perturbations, emission from the SPE undergoes random jumps in its emission energy and its brightness. This is termed as spectral/intensity jitter of the SPE. If the emission turns on and off randomly in time, it is termed as blinking of the SPE. Furthermore, if the emission quenches completely, it is termed as bleaching of the SPE. This jitter can result in poor interference among the consecutively emitted photons, thereby reducing their indistinguishability.

\section*{Why TMDC platform for SPEs?}{\label{Why_TMDCs}}

Single photon nature of the sharp emission lines in the cryogenic photoluminescence (PL) of monolayer WSe\tsub{2} was first demonstrated in the year 2015 by multiple research groups\cite{chakrabortyVoltagecontrolled2015,heSingle2015,koperskiSingle2015,srivastavaOptically2015}. \textcolor{\mcolor}{These emitters, being the localized versions of the free excitons of the monolayers, inherit their high binding energy due to strong out-of-plane confinement and lower effective dielectric constant of the monolayer.} Soon after that, \textcite{kumarStrainInduced2015} experimentally showed the role of strain in deterministic placement of SPEs by suspending WSe\tsub{2} monolayer over etched holes in the substrate. By next year, multiple implementations with electrical excitation of the SPE had been realized\cite{clarkSingle2016,schwarzElectrically2016,palacios-berraqueroAtomically2016}. In 2017, \textcite{palacios-berraqueroLargescale2017} demonstrated an array of SPEs by placing a WSe\tsub{2} monolayer on an array of nanopillars. This method has since been widely adopted as a scalable and deterministic approach to creating bright SPEs\cite{sortinoBright2021,partoDefect2021,brannyDeterministic2017,mukherjeeObservation2020,caiRadiative2018,luoSingle2019,chhaperwalSimultaneously2024}. \textcolor{\mcolor}{Although the majority of the implementations of TMDC-hosted SPEs utilize WSe\tsub{2}, SPEs have been demonstrated in other TMDCs as well such as MoSe\tsub{2}\cite{yuSiteControlled2021}, MoS\tsub{2}\cite{dashQuantum2025,hotgerGateSwitchable2021}, WS\tsub{2}\cite{lohNb2024,cianciSpatially2023}, and MoTe\tsub{2}\cite{zhaoSitecontrolled2021,wyborski_toward_2025}.} A decade later, the field of SPEs in TMDCs continues to witness significant contributions, with rapidly improving figures of merit. This strong research push has sustained because TMDC-based SPEs offer some unique benefits over alternative SPE platforms. We outline some of these benefits below:

\subsection*{Electrical injection}{\label{Electrical injection}}
The ability to trigger single photons on demand by an electrical pulse is desirable for scalable photonic quantum technologies. TMDCs, because of their semiconducting two-dimensional (2D) structure, are much easier to gate compared to quantum dots (QDs), hexagonal boron nitride (hBN), or nanodiamonds. This allows efficient injection of electrons and holes through Fermi-level tuning. These electrons and holes can then form excitons and emit photons upon recombination. Several electrical injection-based SPEs have already been demonstrated with TMDCs \cite{palacios-berraqueroAtomically2016,lenferinkTunable2022,guoElectrically2023,soElectrically2021}. 
\subsection*{Spectral tunability}{\label{Spectral tunability}}
Due to the 2D nature of TMDCs, they can be easily sandwiched in a capacitor-like structure to apply a uniform electrical field in their out-of-plane direction. This electric field can tune the emission energy of defect-based SPEs in TMDCs through the Stark effect \cite{mukherjeeElectric2020}. \revised{Studies have shown that, in addition to ease of fabrication, TMDCs offer superior spectral tunability of $\sim$140-170 meV either via strain or electrical tuning \cite{ciarrocchiPolarization2019, kumarStrainInduced2015}, compared to reported values of $\sim$1-3 meV for QDs \cite{moczala-dusanowska_strain-tunable_2020, schmidt_deterministically_2020, yang_tunable_2024}}. This is desirable because inhomogeneity in the samples often causes the emission energies of distinct SPEs to differ, and the ability to tune their energies can significantly improve the indistinguishability of the SPEs. This also relaxes the need for ultra-high purity of material, as this spectral tuning can overcome slight sample-to-sample inhomogeneity. \\

\subsection*{Deterministic placement of SPEs}{\label{Deterministic placement of SPEs}}
In addition to temporal determinism, spatial determinism is crucial for scalable SPE-based technologies. Defects in hBN and diamond lattices occur randomly, making the location uncontrollable. Self-assembled quantum dots, which are widely used for SPE creation, also lack control over the location of the SPEs. While advancements have been made to address this issue through controlled defect creation in hBN and high-precision placement of Quantum dots, TMDCs offer a much simpler alternative. SPE creation by placing the TMDCs on pre-patterned nanopillar arrays has been shown to work quite reliably \cite{chhaperwalSimultaneously2024,sortinoBright2021,partoDefect2021,brannyDeterministic2017,mukherjeeElectric2020,palacios-berraqueroLargescale2017,caiRadiative2018,luoSingle2019}. This method is easily scalable as chemical vapor deposition (CVD)-grown large-area monolayer TMDCs continue to improve in terms of material quality. The ease of fabricating a large number of SPE centers arranged in a predetermined pattern gives TMDCs a lead in scalable implementations.

\subsection*{Integration with photonic cavities}{\label{Integration with photonic cavities}}
Photonic cavities are believed to play an essential role in SPE-based quantum technologies, offering higher emission rates, shorter lifetimes, and better indistinguishability \cite{flattenMicrocavity2018,Drawer2023-qf,iffPurcellEnhanced2021,azzamPurcell2023,sortinoBright2021,luoDeterministic2018,soPolarization2021,caiRadiative2018,peyskensIntegration2019}. Efficient integration of SPEs with photonic cavities requires precise placement of SPEs within the cavity. The ease of fabricating SPEs at predetermined positions in TMDCs makes their cavity coupling easier \revised{ \cite{mitryakhinEngineering2024} compared to the epitaxial growth of QDs in semiconductors \cite{lodahlInterfacing2015}. Although deterministic coupling of nitrogen vacancy (NV) centers to photonic cavities has been demonstrated recently, the process remains challenging \cite{englund_deterministic_2010}}. The 2D nature of the TMDCs also offers an advantage in this regard, as the presence of a monolayer of TMDC has a negligible effect on the cavity modes.

\subsection*{Higher collection efficiency}{\label{Higher collection efficiency}}
Because of the 2D nature of TMDCs, emitted photons do not suffer a loss from total internal reflection (TIR)\cite{liQuantum2022,partoDefect2021}. Photons emitted in the upper hemisphere by the SPE can travel in the air and be collected directly with minimum loss. On the contrary, SPEs in optical denser media such as diamond, hBN, or quantum dots suffer from TIR-mediated loss, wherein photons emitted even in the upper hemisphere are back-reflected by the material-to-air interface\cite{senellartHighperformance2017,liuBroadband2024,haddenStrongly2010,beveratosRoom2002,wanEfficient2018,lesageEfficient2012}. This results in the actual collected photon rate being significantly lower than the emission rate. This creates a problem for technologies that require time determinism, as photon loss causes the collected photons to be randomly scattered in time. 

\section*{A survey on the origin of SPEs in TMDC}{\label{Origin}}
\textcolor{\mcolor}{For drawing conclusions about the atomic origin of defect centers functioning as SPEs and related quantities such as polarization, fine-structure splitting, g-factor, etc., a large body of literature is ideal. Given the abundance of published material regarding SPEs and defect emission in WSe\tsub{2}, the following subsections focus specifically on this material. But the overall takeaway points are expected to be consistent for other TMDCs as well.}

\subsection*{Atomic origin}{\label{Atomic_Origin}}

Single photon emission from TMDCs requires a highly localized exciton bound to a defect center. The atomic origin of these defect states remains a highly debated topic. Multiple defect complexes have been studied through a wide range of experimental and numerical techniques. Some of the most promising explanations for the origin of these defect states are summarized below:\\\\
\textbf{Selenium vacancy (V\tsub{Se}):} V\tsub{Se} is considered to be the most abundant defect center in WSe\tsub{2} because of its lowest formation energy\cite{zhengPoint2019,Li2019,jeongSpectroscopic2019,wuDefect2017}. V\tsub{Se} breaks the local out-of-plane mirror symmetry as opposed to the double selenium vacancy (V\tsub{SeSe}), which restores it. Stark shift of the emission energy can differentiate between the two, with linear shift corresponding to V\tsub{Se} and quadratic shift corresponding to V\tsub{SeSe}\cite{linhartLocalized2019}.

\begin{figure}[!hbt]
	\centering
	\vs{-0in}
	\hs{-0in}
	\includegraphics[width=0.93\paperwidth,center]{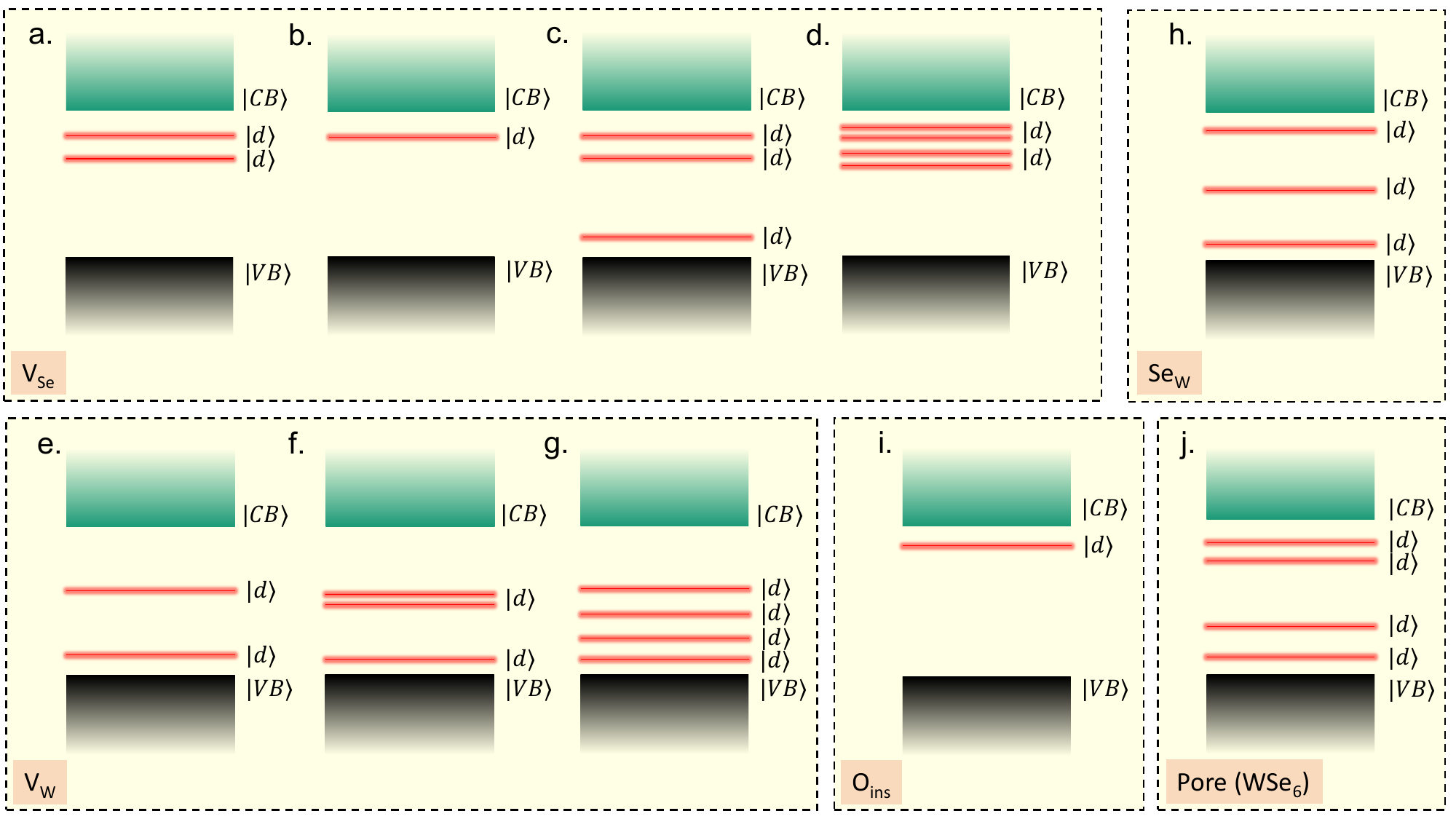}
	\vspace{0in}
	  \caption{\textcolor{\mcolor}{\textbf{Compilation of various assignments of gap states associated with different defect centers as reported in the literature.} \ketm{CB}, \ketm{VB}, and \ketm{d} denote conduction band, valence band, and defect state, respectively. (a-d) Various assignments of the qualitative energy location of the gap states due to V\tsub{Se} center in literature - (a) \cite{partoDefect2021,linhartLocalized2019,wuModulation2025,seratidebritoProbing2024,moodyMicrosecond2018,hernandezlopezStrain2022}, (b) \cite{Li2019,abramovPhotoluminescence2023}, (c) \cite{jeongSpectroscopic2019,qianDefect2020}, (d) \cite{utamaChemomechanical2023}. (e-g) Various assignments of gap states to V\tsub{W} center reported in literature - (e) \cite{zhengPoint2019,Li2019,zhangDefect2017}, (f) \cite{zhangDefect2017,Li2019}, (g) \cite{zhengPoint2019}. (h) Gap states arising from O\tsub{ins} center\cite{jiangTunability2018}. (i) Gap state arising from Se\tsub{W} center\cite{zhengPoint2019}. (j) Gap states arising from pore vacancy (WSe\tsub{6})\cite{partoDefect2021}.}}\label{fig:Defect_states}
\end{figure}

\textcolor{\mcolor}{\autoref{fig:Defect_states}a-d compiles and compares the various assignments of midgap states associated with the V\tsub{Se} center as reported in the literature.} Based on chemomechanical modification of the WSe\tsub{2} monolayer and calculations of the resulting band structure in the presence of V\tsub{Se} center, \textcite{utamaChemomechanical2023} assigned the defect emission from 1.63 eV to 1.67 eV to the V\tsub{Se} defect. States resulting from V\tsub{Se} are proposed to act as single photon emitters upon hybridization \textcolor{\mcolor}{(see \autoref{fig:Role_of_strain}b)} with the dark exciton band of WSe\tsub{2} when the conduction band moves down in energy due to strain\cite{utamaChemomechanical2023,partoDefect2021,wuModulation2025,abramovPhotoluminescence2023,seratidebritoProbing2024,yucelStrain2025}. This explanation is plausible for SPEs 50 - 150 meV below the free exciton emission in monolayer WSe\tsub{2}. Supporting evidence for this mechanism primarily includes the observed spectral distribution of more than 80 SPEs, with a peak at 1.63 eV. This energy position matches the expected spectral position of the V\tsub{Se} defect state as determined through PL imaging combined with atomic force microscopy and strain reconstruction\cite{abramovPhotoluminescence2023}.

Further, theoretical calculations show that V\tsub{Se} yields a singlet $a\tsub{1}$ state near or below the valence band maximum and doublet $e$ states inside the band gap near the conduction band minimum\cite{jeongSpectroscopic2019,moodyMicrosecond2018,linhartLocalized2019,qianDefect2020,nguyenGateTunable2021,partoDefect2021,utamaChemomechanical2023,hernandezlopezStrain2022}. Degeneracy of the doublet $e$ states is expected to be lifted by \simm{}100 - 200 meV due to spin-orbit interaction\cite{jeongSpectroscopic2019}. The upper branch of the doublet, located closer to the dark exciton band, is expected to hybridize with it under strain, resulting in single-photon emission. \textcite{hernandezlopezStrain2022} employed a device architecture with electrostatically tunable strain in monolayer WSe\tsub{2} and found the experimental evidence of two distinct defect states corresponding to V\tsub{Se} as predicted by theory. They determined the emission energies of these two states to be 1.63 eV and 1.45 eV. \textcite{wuModulation2025} performed a similar experiment and found the emission energies of these two states to lie in the 1.65 - 1.67 eV and 1.57 - 1.59 eV range. Another independent evidence of the existence of these two defect states came from photocurrent measurement in a field-effect transistor configuration\cite{nguyenGateTunable2021}.

\textcolor{\mcolor}{Suggestions opposing the role of V\tsub{W} in single-photon emission are also made, with other defect centers proposed as the atomic origin of the SPEs, as discussed below.}\\\\
\textbf{Tungsten vacancy (V\tsub{W}):} V\tsub{W} is proposed to be a dominant point defect in CVD-grown monolayer WSe\tsub{2} based on scanning tunneling microscopy (STM) and spectroscopy (STS) studies\cite{zhangDefect2017}. This vacancy creates a \textit{C\tsub{3}} trigonal symmetry, which is cited to explain conserved valley coherence observed in polarization-resolved magneto-optic measurements\cite{zhangDefect2017,dangIdentifying2020,linhartLocalized2019,heSingle2015}. 

\textcolor{\mcolor}{\autoref{fig:Defect_states}e-g compiles and compares the various assignments of gap states associated with the V\tsub{W} center as reported in the literature.} Theoretically, V\tsub{W} is predicted to create states near the valence band, which can capture holes\cite{Li2019,jiangTunability2018,zhangDefect2017,zhengPoint2019}. \textcite{qianDefect2020} also reported the presence of two hole trapping states near the valence band in addition to three electron trapping states deep in the band gap based on theoretical calculations. Presence of states near the valence band arising from V\tsub{W} has been experimentally confirmed through STM/STS measurements\cite{zhangDefect2017}. \textcite{zhangDefect2017} found that PL from the sample confirmed to have predominantly V\tsub{W} defect centers (through STM) showed a broad emission peak centered at 1.7 eV. \textcite{dangIdentifying2020} assigned one of the quantum emitters in their sample (at \simm{}1.615 eV) as originating from a hole trapped by the V\tsub{W} state near the valence band based on magneto-optic measurements. These assignments contradict the claim of SPEs 50-150 meV below the free exciton emission arising from V\tsub{Se}. \textcite{wuDefect2017} listed V\tsub{W} as one of the possibilities for the broad defect emission (1.3 - 1.5 eV) visible in PL after Ar plasma treatment of the WSe\tsub{2} monolayer.\\\\
\textbf{Antisite defect (Se\tsub{W}):} Antisite defects like Se\tsub{W}, where a selenium atom replaces tungsten in the lattice, are shown to create states both deep in the band gap and near the valence band maximum\cite{zhangDefect2017,zhengPoint2019,Li2019} \textcolor{\mcolor}{(see \autoref{fig:Defect_states}h)}. Calculations by \textcite{zhengPoint2019} suggest a broad range of states down to 300 meV below the exciton energy for the Se\tsub{W} defect. Contrary to this, \textcite{dassUltraLong2019} attributed a sharp emission peak at 1.67 eV with a long lifetime of 225 ns to this defect center based on theoretical calculations. Se\tsub{W} breaks the \textit{C\tsub{3v}} trigonal symmetry created by V\tsub{W}\cite{zhangDefect2017,dassUltraLong2019}, which is cited to explain conserved valley coherence observed in polarization-resolved magneto-optic measurements\cite{zhangDefect2017,dangIdentifying2020,linhartLocalized2019,heSingle2015}.\\\\
\textbf{Oxygen-related defects (O\tsub{Se}, O\tsub{ins}):} In ambient condition, environmental oxygen (O\tsub{2}) can dissociate at the intrinsic defect locations and form oxygen complexes such as O\tsub{2} bound at V\tsub{SeSe} site, O bound at V\tsub{Se} site (O\tsub{Se}), O occupying interstitial site (O\tsub{ins}), and O adsorbed on Se (O\tsub{ad})\cite{zhengPoint2019}. \textcite{zhengPoint2019} imaged CVD grown monolayer WSe\tsub{2} in STM and identified three distinct defect centers D1, D2, and D3. None of these showed any gap states in STS or single-particle calculations, which the intrinsic defects are predicted to create, as discussed in the previous sections. Furthermore, these defects D1, D2, and D3 matched well with simulated STM images for O\tsub{Se}, O\tsub{ins}, and O\tsub{ad} respectively. Other studies have also shown that oxygen-passivated selenium vacancies do not have a mid-gap state in the single-particle picture\cite{partoDefect2021}. But for multi-particle calculations (GW-BSE calculation with the perturbative spin-orbit coupling), O\tsub{ins} showed a localized exciton state LX\tsub{1} 84 meV below A exciton \textcolor{\mcolor}{(see \autoref{fig:Defect_states}i)}, and two more LX\tsub{2} and LX\tsub{3} localized states above A exciton of the monolayer WSe\tsub{2}\cite{zhengPoint2019}. Since the capture process by the defect requires a defect state with lower energy than the free exciton, the LX\tsub{1} state is a promising candidate for single photon emission, especially for samples in direct contact with ambient oxygen. LX\tsub{1} state also lies in the same range as the relatively higher energy SPEs (45-100 meV below the A exciton) observed in literature. O\tsub{Se} did not show any state below the A exciton in the multi-particle calculations\cite{zhengPoint2019}.\\\\
\textbf{Pore vacancies:} Pore vacancies such as WSe\tsub{6} complex are proposed as the origin of sharp single photon emission peaks around 1.5 - 1.55 eV\cite{partoDefect2021}. This range is approximately 100 meV deeper than the usual SPE spectral range. \textcite{partoDefect2021} calculated the energy dynamics of WSe\tsub{6} pore with varying strain and showed that these complexes not only create states below the conduction band \textcolor{\mcolor}{(see \autoref{fig:Defect_states}j)}, but can hybridize with the band as well. These states are at lower energy than the ones formed by V\tsub{Se} defects and thus can support another class of SPEs with emission energy in the 1.5 - 1.55 eV range. \\

% ------------------------------------------------------
\subsection*{Role of strain}{\label{Role_of_strain}}
Application of tensile strain on monolayer WSe\tsub{2} has been shown to lower its conduction band minimum (CBM) both in theoretical calculations\cite{feierabendDark2019,partoDefect2021,Chang2013,hernandezlopezStrain2022,aminStrain2014,lvStrainDependent2023,desaiStrainInduced2014,Brooks2018,linhartLocalized2019} as well as in experiment\cite{Shen2016a}. But the movement of the valence band maximum (VBM) is not well agreed upon. Support for all three cases - VBM moving up in energy\cite{Chang2013}, VBM moving down in energy\cite{feierabendDark2019,hernandezlopezStrain2022,linhartLocalized2019}, and VBM being independent of tensile strain\cite{lvStrainDependent2023,desaiStrainInduced2014,Shen2016a} - is present in the literature. However, both upward and downward movements of VBM have been shown to be significantly weaker compared to the conduction band movement. This is clear when comparing the quantitative rate of shift of CBM/VBM per \% of tensile strain in monolayer WSe\tsub{2} as summarized in \autoref{tab:bandmovementwithstrain}. This results in the conclusion of reduction of bandgap with tensile strain which has been confirmed both theoretically\cite{partoDefect2021,Chang2013,hernandezlopezStrain2022,frisendaBiaxial2017,defoStrain2016,lvStrainDependent2023,desaiStrainInduced2014,Brooks2018,Johari2012,feierabendDark2019,linhartLocalized2019,schmidtReversible2016,aminStrain2014,kumarStrain2024,zollnerStraintunable2019} and experimentally\cite{frisendaBiaxial2017,schmidtReversible2016,Shen2016a,kumarStrain2024,aslanStrain2018}. The rate of change in bandgap per \% of tensile strain is also summarized in \autoref{tab:bandmovementwithstrain}.
\renewcommand{\arraystretch}{1.3}
\begin{table}
    \centering
    \begin{tabular}{|c|p{2.3cm}|c|c|c|c|}
        \hline
        Strain type & Theoretical/\newline Experimental & CBM & VBM & E\tsub{g} & Ref.\\[1.1ex]
        \hline
        Biaxial & Theoretical & -140 meV/\% & -40 meV/\% & -100meV/\% & \cite{linhartLocalized2019}\\
        \hline
        Biaxial & Theoretical & $-$ & $-$ & -134 meV/\% & \cite{frisendaBiaxial2017}\\
        \hline
        Biaxial & Theoretical & Down$^*$ & Down$^*$ & -135 meV/\% & \cite{feierabendDark2019}\\
        \hline
        Biaxial & Theoretical & Down$^*$ & $-$ & -95 meV/\% & \cite{aminStrain2014}\\
        \hline
        Biaxial & Theoretical & $-$ & $-$ & -118.8 meV/\% & \cite{kumarStrain2024}\\
        \hline
        Biaxial & Theoretical & $-$ & $-$ & -121.1 meV/\% & \cite{zollnerStraintunable2019}\\
        \hline
        Uniaxial & Theoretical & $-$ & $-$ & -43 meV/\% & \cite{schmidtReversible2016}\\
        \hline
        Biaxial & Experimental & $-$ & $-$ & -63 meV/\% & \cite{frisendaBiaxial2017}\\
        \hline
        Biaxial & Experimental & $-$ & $-$ & -118 meV/\% & \cite{kumarStrain2024}\\
        \hline
        Uniaxial & Experimental & $-$ & $-$ & -47.6 meV/\% & \cite{aslanStrain2018}\\
        \hline
        Uniaxial & Experimental & $-$ & $-$ & -54 meV/\% & \cite{schmidtReversible2016}\\
        \hline
        Uniaxial & Experimental & -50 to -100 meV/\% & 0 meV/\% & -50 to -100 meV/\% & \cite{Shen2016a}\\
        \hline
        Biaxial & Theoretical & Down$^*$ & $-$ & Down$^*$ & \cite{partoDefect2021}\\
        \hline
        Biaxial & Theoretical & Down$^*$ & Down$^*$ & Down$^*$ & \cite{hernandezlopezStrain2022}\\
        \hline
        Biaxial & Theoretical & Down$^*$ & Up$^*$ & Down$^*$ & \cite{Chang2013}\\
        \hline
        Uniaxial & Theoretical & Down$^*$ & Up$^*$ & Down$^*$ & \cite{Chang2013}\\
        \hline
        Biaxial & Theoretical & Down$^*$ & Fixed$^*$ & Down$^*$ & \cite{lvStrainDependent2023}\\
        \hline
        Biaxial & Theoretical & Down$^*$ & Fixed$^*$ & Down$^*$ & \cite{desaiStrainInduced2014}\\
        \hline
        Biaxial & Theoretical & Down$^*$ & $-$ & Down$^*$ & \cite{Brooks2018}\\
        \hline
    \end{tabular}
    \caption{Movement of conduction band minimum (CBM), valence band maximum (VBM), and band gap (E\tsub{g}) of WSe\tsub{2} under tensile strain.\\$^*$Quantitative values not available.}
    \label{tab:bandmovementwithstrain}
\end{table}

This reduction of bandgap/CBM energy can influence the single photon emission in two ways, as summarized below:
\subsubsection*{Funnelling of excitons}
\begin{figure}[!hbt]
	\centering
	\vs{-0.3in}
	\hs{-0in}
	\includegraphics[width=0.9\paperwidth,center]{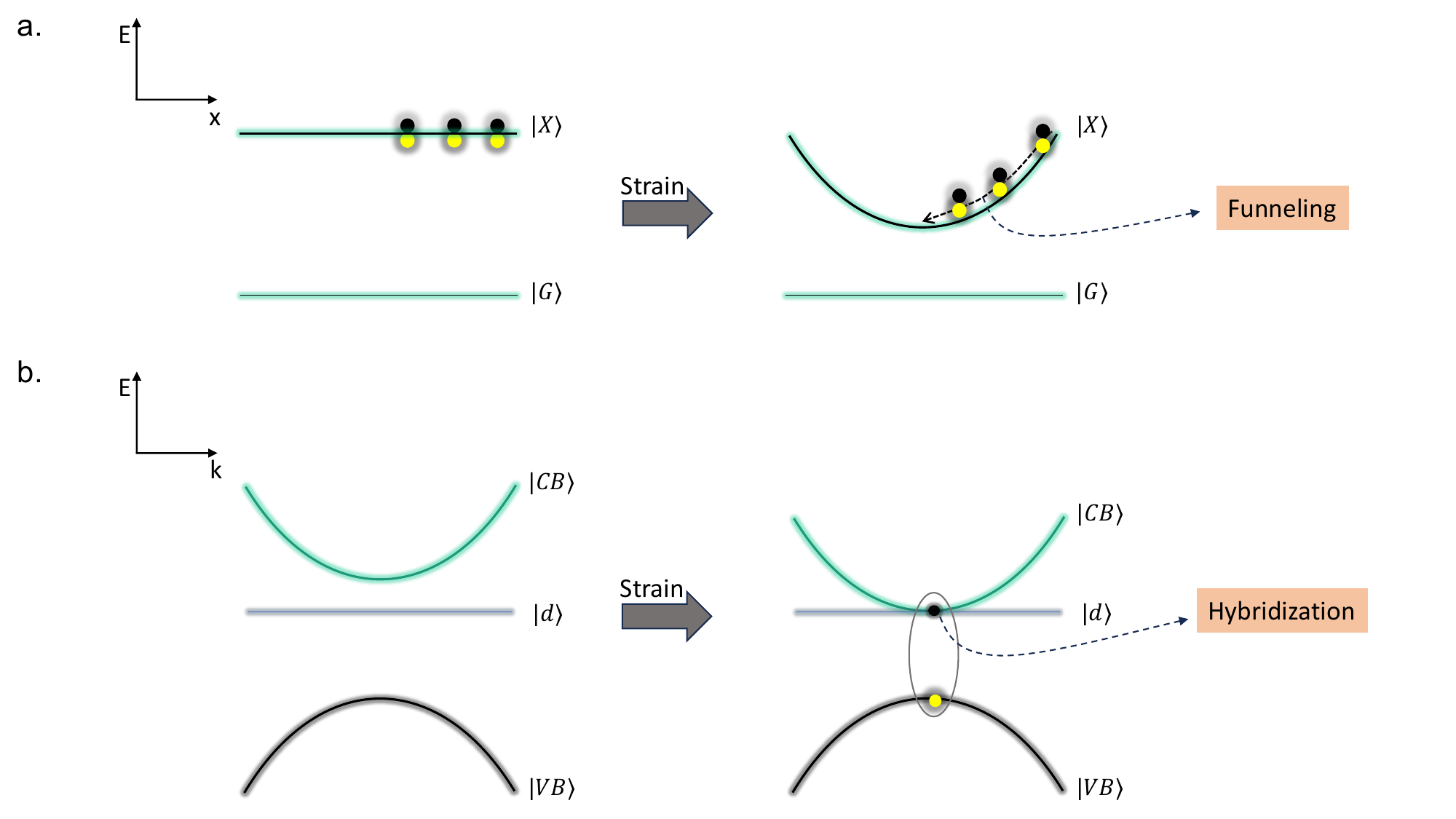}
	\vspace{-0in}
	  \caption{\textbf{Role of tensile strain in relation to SPEs.} (a) Tensile strain reduces the bandgap of WSe\tsub{2} monolayer proportional to its magnitude, resulting in an energy well for the excitons. Excitons from other parts of the sample then funnel to the strain-induced energy minimum. (b) Conduction band of WSe\tsub{2} monolayer moves down in energy with tensile strain, resulting in its hybridization with the defect state located energetically close to it.}\label{fig:Role_of_strain}
\end{figure}

Since the bandgap reduction is proportional to the magnitude of tensile strain applied (\autoref{tab:bandmovementwithstrain}), non-uniform strain causes the bandgap to shift inhomogeneously in space (\autoref{fig:Role_of_strain}a)\cite{Brooks2018}. This is the case when a monolayer of WSe\tsub{2} is placed on nanopillars or other strain-inducing structures. The top of the nanopillar, being the most strained, is the region of the bandgap minimum, which causes excitons from surrounding locations to funnel to this location of minimum energy (\autoref{fig:Role_of_strain}a)\cite{brannyDeterministic2017,chhaperwalSimultaneously2024}. A defect center, if present at this location, receives a large supply of excitons to capture and subsequently recombine to emit single photons. This is the basis of multiple implementations of SPEs with nanopillar arrays\cite{sortinoBright2021,partoDefect2021,brannyDeterministic2017,mukherjeeObservation2020,palacios-berraqueroLargescale2017,caiRadiative2018,luoSingle2019,chhaperwalSimultaneously2024}. PL map of such samples with a nanopillar array shows predominant emission of excitons from the pillar centers compared to the surrounding region, confirming the role of strain in exciton funnelling\cite{palacios-berraqueroLargescale2017,chhaperwalSimultaneously2024}.
\subsubsection*{Hybridization of defect state with conduction band} As discussed, the conduction band of monolayer WSe\tsub{2} moves down in energy with tensile strain\cite{feierabendDark2019,partoDefect2021,Chang2013,hernandezlopezStrain2022,aminStrain2014,lvStrainDependent2023,desaiStrainInduced2014,Brooks2018,linhartLocalized2019,Shen2016a}. The effect of this strain on the energy of the defect state is comparatively weaker\cite{hernandezlopezStrain2022,yuDynamic2025,iffStrainTunable2019}. This means that with sufficient strain, the lower conduction band can come energetically very close to the mid-gap defect states, resulting in hybridization between them (\autoref{fig:Role_of_strain}b)\cite{utamaChemomechanical2023,partoDefect2021,wuModulation2025,abramovPhotoluminescence2023,seratidebritoProbing2024,yucelStrain2025}. This hybridization has been demonstrated in theoretical calculations involving mid-gap states created due to V\tsub{Se} defect center\cite{partoDefect2021,linhartLocalized2019}. V\tsub{Se} defect centers are generally considered for this scenario because these are considered to be the most abundant in WSe\tsub{2}\cite{zhengPoint2019,Li2019,jeongSpectroscopic2019,wuDefect2017}, and because the resulting mid-gap states are close enough to the conduction band for strain to hybridize them. Such hybridization can brighten the dark intervalley excitons - with electron in the lower conduction band of WSe\tsub{2} (\autoref{fig:Role_of_strain}b)\cite{linhartLocalized2019}. \textcite{hernandezlopezStrain2022} employed an architecture with electrostatically tunable tensile strain and demonstrated this hybridization at two energy positions - 1.63 eV and 1.45 eV - in line with the prediction of two defect states created by the V\tsub{Se} defect center. In a similar experiment, \textcite{wuModulation2025} demonstrated such hybridization for two energy ranges - 1.65-1.67 eV and 1.57-1.59 eV - again in line with the prediction of two defect states created by the V\tsub{Se} defect center. \textcite{abramovPhotoluminescence2023} tracked the position of the dark exciton band with strain and found that the SPE emission energy closely follows it, proving the hybridization of the conduction band with different defect states based on the strain.  \textcite{partoDefect2021} claimed that V\tsub{Se} centers can easily be passivated with oxygen and thus may not be the source of SPEs observed. They repeated the calculations for WSe\tsub{6} pore vacancy complex and demonstrated that, like defect states of V\tsub{Se}, defect states from this defect center can also hybridize with the conduction band in the presence of tensile strain.

These two effects are not mutually exclusive and are repeatedly cited together to explain the SPEs in deterministically strained WSe\tsub{2}\cite{xuConversion2023,yuDynamic2025,linhartLocalized2019,hernandezlopezStrain2022}.

\subsection*{Fine structure splitting, g-factor, and origin of polarization}

{\label{FSS_g-factor_Pol}}
Observed PL emission peak from defect-based single photon emitters in WSe\tsub{2} fall into two broad categories:
\subsubsection*{Fine structure split doublet peaks}
\begin{figure}[!hbt]
	\centering
	\vs{-0.3in}
	\hs{-0in}
	\includegraphics[width=0.95\paperwidth,center]{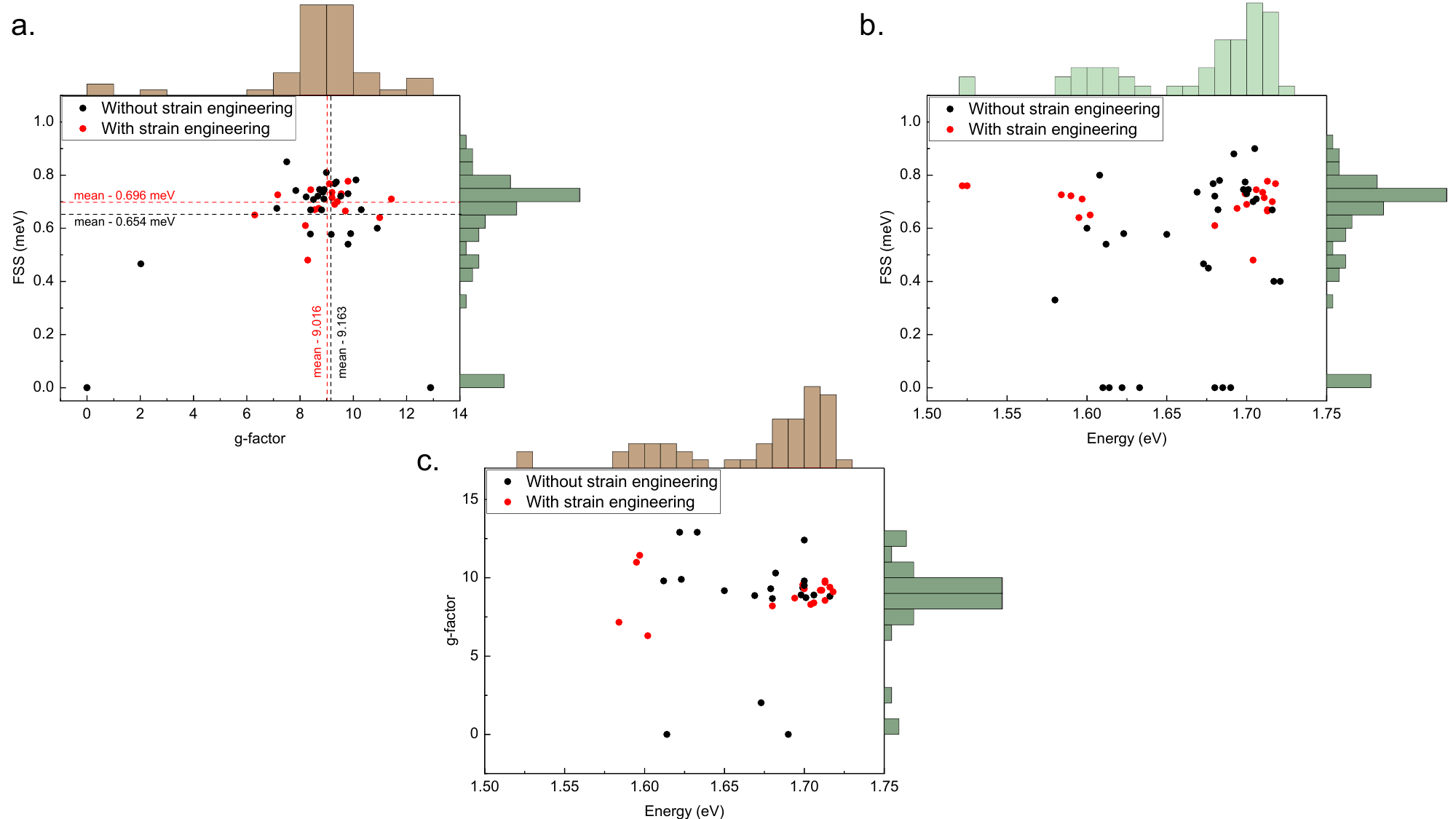}
	\vspace{-0in}
	  \caption{\textbf{Reported values of fine structure splitting and g-factor for defect emitters at different energies in the literature.} (a) Reported values of FSS are plotted against the respective g-factor values for multiple emitters in literature. Data from emitters with (without) strain engineering is shown in red (black) circles. The average value is calculated excluding zero values of FSS and g-factor. Histograms of the FSS and g-factor values are shown at the top and right of the plot, respectively. (b) Reported values of FSS are plotted against the emission energies for multiple emitters in the literature.  Histograms of the FSS values and emission energies are shown at the top and right of the plot, respectively. (c) Reported values of the g-factor are plotted against the emission energies for multiple emitters in the literature.  Histograms of the g-factor values and emission energies are shown at the top and right of the plot, respectively. In all the plots, data from emitters with (without) strain engineering are shown in red (black) circles.}\label{fig:fss_g-factor}
\end{figure}

\begin{figure}[!hbt]
	\centering
	\vs{-0.3in}
	\hs{-0in}
	\includegraphics[width=0.9\paperwidth,center]{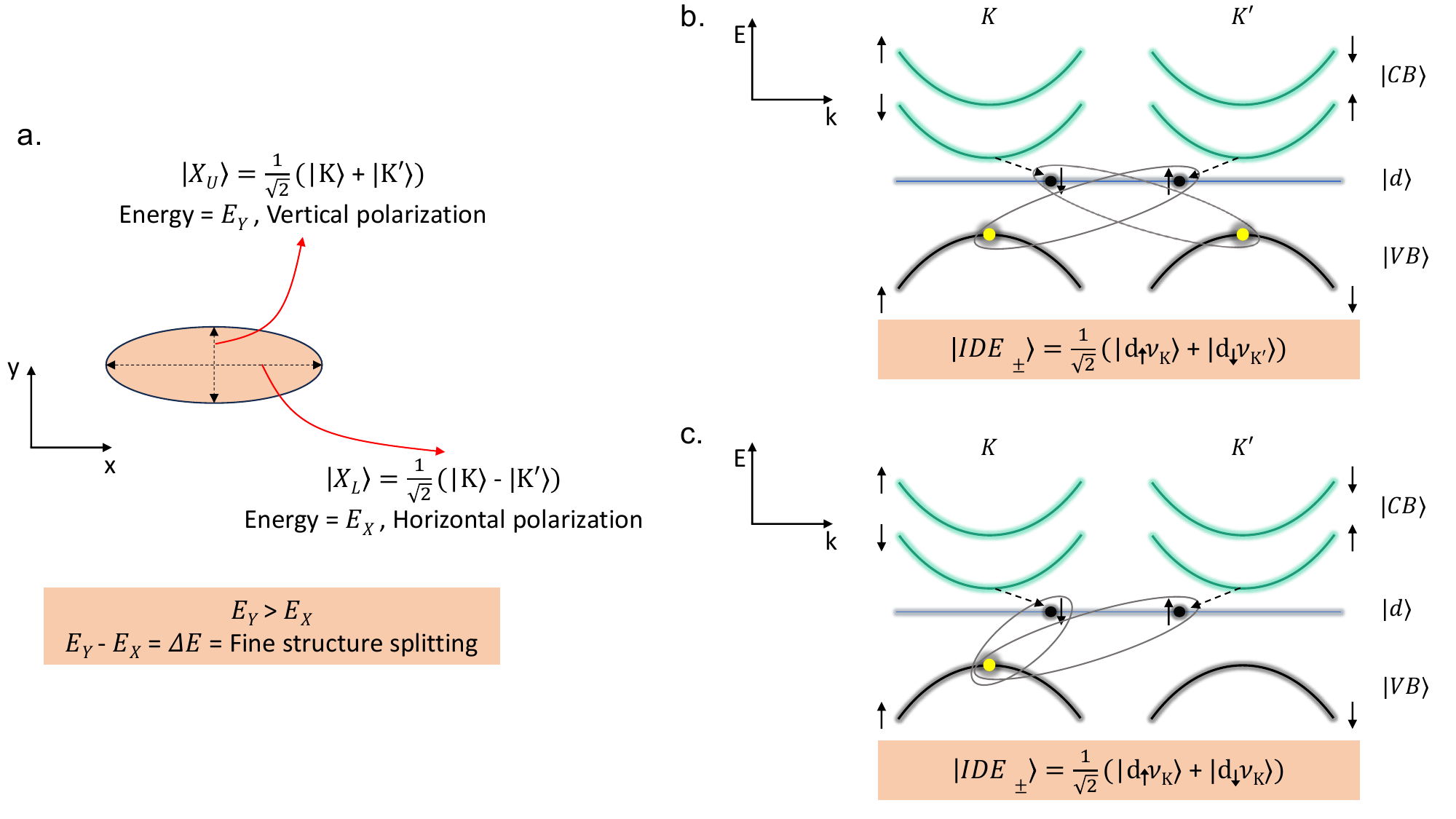}
	\vspace{-0in}
	  \caption{\textbf{Fine structure splitting as a result of either anisotropic confinement or intervalley mixing.} (a) An elliptical confinement potential is shown, which breaks the in-plane x-y symmetry and splits the exciton peaks into two orthogonal linearly polarized peaks. Each of the two states can be written as the superposition of $K$ and $K'$ valley excitons\cite{wangHighly2021}. (b) SPE considered as a superposition of two intervalley defect excitons. Dashed arrow shows the source of electrons (black balls) of each spin. Each of the electrons forms an exciton with a hole (yellow balls) of the same spin\cite{linhartLocalized2019}. (c) SPE considered as a superposition of one intravalley and one intervalley exciton\cite{xiangMagnetic2025}.}\label{fig:ellipse_and_intervalley}
\end{figure}

General observation for the quantum emitters in WSe\tsub{2} indicates two split peaks in PL with an energy gap of \simm{}0.7 meV between them\cite{chakraborty3D2018,heCascaded2016,chakrabortyElectrical2019,wangHighly2021,dangIdentifying2020,srivastavaOptically2015,heSingle2015,debritoStrain2022,chakrabortyVoltagecontrolled2015,luOptical2019,luoDeterministic2018}. The value of this splitting lies remarkably close for most of the emitters studied in the literature (with some exceptions, discussed later), as shown in \autoref{fig:fss_g-factor}a,b. These two peaks are cross-linearly polarized in most cases \cite{heCascaded2016,partoDefect2021,wangHighly2021,dangIdentifying2020,linhartLocalized2019,xiangMagnetic2025,seratidebritoProbing2024,yangRevealing2023,clarkSingle2016,heSingle2015,debritoStrain2022,luOptical2019}. The splitting between the peaks increases under an out-of-plane magnetic field, leading to a g-factor of \simm{}9\cite{chakraborty3D2018,wangHighly2021,srivastavaOptically2015,seratidebritoProbing2024,heSingle2015,debritoStrain2022}. The value of the g-factor also lies in a narrow range around the mean for most of the emitters (with some exceptions, discussed later) as shown in \autoref{fig:fss_g-factor}a,c. As the magnetic field intensity increases, the splitting increases and the peaks begin to exhibit circular dichroism\cite{chakrabortyElectrical2019,wangHighly2021,dangIdentifying2020,srivastavaOptically2015,seratidebritoProbing2024,heSingle2015,debritoStrain2022,luOptical2019,koperskiSingle2015}. There have been significant attempts to explain all of these observations under a unified mechanism, but the debate is still ongoing. Below, we categorize these explanations into two broad classes:\\\\
\textbf{Structural anisotropy:} This explanation relies on symmetry breaking of the confinement potential of the defect-bound exciton. This anisotropy can stem from either the defect structure itself or from the strain profile. The exciton is thought to be bound in such cases in a potential with $<$D\tsub{2d} symmetry\cite{chakrabortyVoltagecontrolled2015,wangHighly2021}, which can be modeled as an ellipse for simplicity\cite{wangHighly2021}. Due to broken x-y symmetry, the bound exciton can have its oscillating dipole moment along either the major or the minor axis of the elliptical potential, as shown in \autoref{fig:ellipse_and_intervalley}a. The electron-hole wavefunction overlap in these two configurations differs, leading to an anisotropic exchange interaction. This results in the lifting of degeneracy between the two configurations, manifesting as the fine structure split between two peaks in the PL spectrum\cite{chakraborty3D2018,heCascaded2016,chakrabortyElectrical2019,wangHighly2021,dangIdentifying2020,srivastavaOptically2015,heSingle2015,debritoStrain2022,chakrabortyVoltagecontrolled2015,luOptical2019,luoDeterministic2018}. This explanation originates from the field of quantum dots, where asymmetric dots have been shown to exhibit a fine split in the energy of the emission peak, albeit to a lesser degree due to a weaker electron-hole interaction\cite{srivastavaOptically2015,chakrabortyElectrical2019,heSingle2015,bayerFine2002,gammonHomogeneous1996}.

Cross-linear polarization can be easily explained within this framework, with two oscillating dipole directions directly corresponding to the two orthogonal polarization directions\cite{wangHighly2021}. Strong evidence for linear polarization emerging from anisotropic confinement is provided by a study on the 1D confinement of excitons in WSe\tsub{2}, which shows that the polarization direction precisely matches the direction of the 1D confinement\cite{wangHighly2021}. \textcite{paralikisTailoring2024} also demonstrated this by fabricating nanowrinkles in the WSe\tsub{2} monolayer in a deterministic direction and matching them with the polarization direction. \textcite{chakrabortyElectrical2019} proved that the two peaks result from two differently oriented dipoles by showing different Stark shifts for both of these in an electric field. Since the two peaks shift at different rates, the authors were able to tune the fine structure between them through the electric field. By approaching a vanishing fine structure split, they achieved a degree of circular polarization (DOCP) of 42\% compared to the DOCP of 0\% without an electric field.

The valley structure of the free WSe\tsub{2} exciton is considered to be inherited by the defect-bound exciton\cite{luOptical2019}, which explains the movement of valleys in out-of-plane magnetic field and thus the Zeeman shift of the doublet energies in the opposite direction (increasing splitting)\cite{luOptical2019}. This inheritance is supported by STM measurements, which have shown trigonal symmetry of the defect states for a certain type of vacancies\cite{zhangDefect2017}.

Explanation of the observed g factor of \simm{}8-10 is difficult in this framework since the free exciton from which the defect-bound exciton is supposed to borrow the valley structure shows a nominal g-factor of \simm{}4\cite{srivastavaValley2015,wangMagnetooptics2015,stierMagnetooptics2018,koperskiOrbital2018,liMomentumDark2019,cianciSpatially2023}. To mitigate this, hybridization of the lower conduction band of WSe\tsub{2} (which moves down in energy with strain\cite{feierabendDark2019,partoDefect2021,Chang2013,hernandezlopezStrain2022,aminStrain2014,lvStrainDependent2023,desaiStrainInduced2014,Brooks2018,linhartLocalized2019,Shen2016a}) with the defect state is suggested. This could explain the observed high g-factor since the spin dark exciton in WSe\tsub{2} (coming from the lower conduction band) has experimentally shown a g factor of \simm{}8\cite{liuGate2019,robertFine2017,liMomentumDark2019,bremPhononAssisted2020}. Building on the valley physics of the free exciton, the defect-bound exciton trapped in an anisotropic potential is suggested to have an equal mix of $K$ and $K^\prime$ intravalley excitons, and the state can be written as\cite{wangHighly2021}:
\begin{align}
\ketm{X\tsub{L}} = \alpham{}\ketm{K} - \betam{}\ketm{K'}\\
\ketm{X\tsub{U}} = \alpham{}\ketm{K} + \betam{}\ketm{K'}
\end{align}
where, \ketm{X\tsub{L(U)}} is the state with lower (higher) energy, $|\alpham{}|\tsup{2}+|\betam{}|\tsup{2}=1$, and \alpham{} = \betam{} in absence of magnetic field. An out-of-plane magnetic field forces one of the valleys to have a lower energy, resulting in mixing of the two valleys in unequal proportions (\alpham{} \neqm{} \betam{}). This tilts the mixture towards elliptical polarization and thus circular polarization is restored at elevated magnetic fields\cite{wangHighly2021,dangIdentifying2020,srivastavaOptically2015}.\\\\
\textbf{Intervalley mixing:} Another major class of formalism is hybridization between intervalley excitons. This explanation accounts for the high values of the g-factor observed, even without considering the conduction band-defect state hybridization (although it can be included to explain the distribution of g-factor values around the mean)\cite{linhartLocalized2019}. This is because while the valley orbital moment cancels out for intravalley excitons (resulting in a low g-factor of \simm{}4\cite{srivastavaValley2015,wangMagnetooptics2015,stierMagnetooptics2018,koperskiOrbital2018,liMomentumDark2019,cianciSpatially2023}), it adds up for the intervalley excitons, resulting in a g-factor of \simm{}13 for momentum dark free excitons in WSe\tsub{2}\cite{forsteExciton2020,liMomentumDark2019,bremPhononAssisted2020}.

The high brightness of the defect-bound emitters in this seemingly momentum-dark configuration stems from the fact that the defect state, being highly localized in real space, does not possess a specific momentum. In other words, the valley selection rules of the free exciton do not apply to the defect-bound excitons\cite{linhartLocalized2019,xiangMagnetic2025}.

In this formalism, the defect state does not inherit the valley physics of the free exciton, as opposed to the previous formalism. \textcite{yangRevealing2023} integrated quantum emitters in WSe\tsub{2} with chiral cavities to prove this lack of spin-valley locking. They demonstrated that the emission from the defect state followed the circular polarization of the chiral cavity, rather than the supposed valley structure.

A mathematical formalism for this mixture of intervalley defect states comes from \textcite{linhartLocalized2019}. They wrote the expression for such a superposition as a symmetric and antisymmetric sum of two transitions: up (down) spin electron [coming from the lower conduction band of the $K'$ ($K$) valley] trapped in the defect state recombining with an up (down) spin hole in the $K$ ($K^\prime$) valley:
\begin{align}
\ketm{IDE\tsub{\pmm{}}} \coloneq \fracm{1}{\sqrtm{2}}(\ketm{d_{\uparrow}v\tsub{K}} \pmm{} \ketm{d_{\downarrow}v\tsub{K$^\prime$}})
\end{align}
where $d_{\uparrow (\downarrow)}$ represents an electron with up (down) spin in the defect state, and $v\tsub{K (K$^\prime$)}$ represents a hole in the $K$ ($K'$) valley. This configuration is schematically shown in \autoref{fig:ellipse_and_intervalley}b. Both of these transitions are bright due to the defect center, which relaxes the valley selection rules. These two states exhibit different electron-hole exchange interactions, based on the \pmm{} sign, resulting in a fine structure splitting.

This mixing of opposite chiralities naturally results in cross-linear polarization of the two emission peaks. But this formalism, independent of the physical structure of the confinement potential, can not predict the direction of this linear polarization with respect to the sample geometry\cite{linhartLocalized2019}. Mixing of the opposite chiralities is suppressed in the out-of-plane magnetic field, returning the circular polarization of the two distinct intervalley excitons. \textcite{seratidebritoProbing2024} claimed that the observed anomalous PL energy redshift under a parallel magnetic field in their sample is evidence of this intervalley nature. Bringing the conduction band-defect state hybridization in this picture results in further splitting of the defect state, possibly manifesting as a correlated couple of doublet emission peaks. A variation of this formalism is presented by \textcite{xiangMagnetic2025}, where they construct the superposition state from these two transitions: up (down) spin electron [coming from the lower conduction band of the $K'$ ($K$) valley] trapped in the defect state recombining with an up spin hole in the $K$ valley:
\begin{align}
\ketm{IDE\tsub{\pmm{}}} \coloneq \fracm{1}{\sqrtm{2}}(\ketm{d_{\uparrow}v\tsub{K}} \pmm{} \ketm{d_{\downarrow}v\tsub{K}})
\end{align}
The first term here corresponds to a bright exciton (no longer momentum dark due to the defect center relaxing the valley selection rules), and the second term corresponds to a spin dark exciton. This configuration is schematically shown in \autoref{fig:ellipse_and_intervalley}c. Application of an in-plane magnetic field brightened the defect emission, which the authors claimed is evidence of field-induced state mixing between bright and dark intravalley excitons\cite{xiangMagnetic2025}.\\

\subsubsection*{Singlet peak}
In contrast to the fine-structure split peaks discussed above, some emitters have been shown to exhibit only a single peak in PL emission\cite{dangIdentifying2020,heSingle2015,kumarResonant2016,wangHighly2021,luOptical2019,chakrabortyVoltagecontrolled2015}. \textcite{heSingle2015} mentioned that such singlet peaks contribute to 40\% of all the peaks observed in their study. These peaks convert from linearly polarized emission to circularly polarized emission with an out-of-plane magnetic field\cite{dangIdentifying2020,wangHighly2021,luOptical2019} just like the doublet peaks. Multiple explanations for these are present in the literature as summarized below:\\\\
\textbf{Charged defect-bound exciton:} Presence of a third particle in the complex can suppress the exchange interaction by forming a singlet pair with the other identical particle\cite{luOptical2019,chakraborty3D2018}. Since the fine structure splitting stems from the exchange interaction according to literature\cite{chakraborty3D2018,heCascaded2016,chakrabortyElectrical2019,wangHighly2021,dangIdentifying2020,srivastavaOptically2015,heSingle2015,debritoStrain2022,chakrabortyVoltagecontrolled2015,luOptical2019,luoDeterministic2018}, a charged defect-bound exciton with three particles can partially or completely suppress the splitting between the two peaks. \textcite{luOptical2019} identified a pair of a doublet peak and a singlet peak arising around the same gate voltage in the p-doping regime in their gate voltage-dependent PL. They attributed the singlet peak to a defect-bound positive trion with no fine structure splitting. They argued that intervalley exchange is suppressed because the two holes form a singlet pair, and intravalley exchange is suppressed due to Pauli blocking of hole exchange. The Pauli blocking comes into play due to the large SOC split in the valence band, which forces both of the holes to occupy the upper valence band. This is not the case for the two electrons in a negative trion, since the SOC split in the conduction band is relatively lower. Therefore, fine structure splitting is lower but not completely suppressed in the case of a negative trion as observed by \textcite{chakraborty3D2018}. The contribution of hole(s) in the valence band (predominant contributor\cite{dangIdentifying2020}) to the g-factor is similar for trions and neutral excitons. Accordingly, the g-factor of these positive and negative defect-bound trions is measured to be 12.9\cite{luOptical2019} and \simm{}8\cite{chakraborty3D2018}, respectively, in a similar range to the neutral defect-bound excitons.\\\\
\textbf{Suppression of higher energy peak:} This is the case where the observation of a single peak does not necessarily correspond to the absence of the doublet energy states. \textcite{wangHighly2021} claimed that for structures with high anisotropy, such as the sharp edge of a cube in their structure, most of the exciton population relaxes to the lower energy state before emission. This results in a dramatic reduction in the emission from the higher energy branch of the doublet, leading to the observation of a single peak in the PL. Since these are actually originating from the doublet states, their g-factor values are found to be similar to the normally observed doublets as well\cite{wangHighly2021}.\\

There are some notable exceptions to the general observations discussed above. In contrast to the closely matching fine structure splitting values (\simm{}0.7 meV) observed for most of the emitters, some outliers include values of 0.4 meV\cite{heCascaded2016}, 0.466 meV\cite{dangIdentifying2020}, 0.45 meV\cite{dassUltraLong2019}, and 0.33 meV\cite{kumarResonant2016}. The latter two of these also showed co-linearly polarized doublet peaks instead of the usual cross-linear polarization. Doublet peaks with an angle of \simm{}40 - 60\degreem{} with each other have also been observed\cite{kumarStrainInduced2015,chakraborty3D2018}. The outliers in the g-factor measurements include emitter with values 0\cite{dangIdentifying2020,heSingle2015} and 2.02\cite{dangIdentifying2020}.

% ----------------------------------------------------
\subsection*{Gate voltage dependence}{\label{Gate_voltage}}

The dependence of the SPEs in TMDCs on their electrostatic environment has yet to be sufficiently studied. A few studies that used gate voltage to tune the properties of the SPEs demonstrated contrasting results. A hole-doped regime was favored for some of these emitters\cite{stevensEnhancing2022,chenEntanglement2019,luOptical2019}, while an electron-doped regime was favored for others\cite{chakraborty3D2018,paralikisTunable2025}. Some studies also concluded that a neutral doping regime was optimal\cite{chakraborty3D2018,chakrabortyVoltagecontrolled2015}. Explanations for these observation include: (1) the emitter being a negative defect-bound trion (with lower energy than the defect-bound neutral exciton)\cite{chakraborty3D2018}, (2) the emitter being a positive defect-bound trion (with higher energy than the defect-bound neutral exciton)\cite{luOptical2019}, (3) negative gate voltage driving more carriers into the strain induced potential well thereby increasing the emission rate\cite{stevensEnhancing2022}, and (4) negative gate voltage leading to a charge imbalance causing the emission to quench\cite{paralikisTunable2025}.
\subsection*{Temperature dependence}{\label{TEmp_SPE}}
The effect of increasing sample temperature on single photon emitters has been studied for two parameters: their linewidth and emission intensity. Both of these are summarized below:
\subsubsection*{Effect of temperature on SPE linewidth}
Linewidth of defect-based SPEs in WSe\tsub{2} is shown to broaden with increasing temperature. This has been attributed to phonon-induced pure dephasing, which increases with the population of acoustic and optical phonons in the system\cite{heCascaded2016,partoDefect2021,hePhonon2016,luoSingle2019}. This behaviour is captured using the equation given by
\begin{align}
\gammam{}(T) = \gammam{}\tsub{0} + \gammam{}\tsub{ac} + \fracm{\gammam{}\tsub{LO}}{e^{\left(\fracm{E_{LO}}{k_{B}T}\right)} - 1}
\end{align}
where, \gammam{}($T$) is the linewidth of the SPE at temperature $T$, $\gammam{}\tsub{0}$ is the projected linewidth of the SPE at 0 K, $k\tsub{B}$ is the Boltzmann constant, \gammam{}\tsub{ac} (\gammam{}\tsub{LO}) is the coefficient of scattering with acoustic (optical) phonons, and $E\tsub{LO}$ is the energy of the optical phonons\cite{heCascaded2016,partoDefect2021,hePhonon2016,luoSingle2019}. It is this dephasing that causes the significant loss of coherence of the SPEs in WSe\tsub{2}\cite{vonhelversenTemperature2023}.

\subsubsection*{Effect of temperature on SPE emission intensity}
A common observation in the literature is that defect-based SPEs primarily function at cryogenic temperatures, with their emission intensity dropping sharply as the temperature increases. This limits the SPE operation to temperatures under \simm{}35 K\cite{srivastavaOptically2015,chakrabortyVoltagecontrolled2015,kumarStrainInduced2015,tonndorfSinglephoton2015,heSingle2015,kernNanoscale2016,heCascaded2016,brannyDeterministic2017,palacios-berraqueroLargescale2017,luoDeterministic2018,kimPosition2019}. The unanimous explanation for this quenching is excitation of the defect-bound exciton out of the localized potential through phonon interaction. Since the population of phonons in the system follows Bose-Einstein statistics, it increases exponentially with temperature. The SPE intensity versus temperature plot can thus be fitted with the Arrhenius equation given by
\begin{align}
I(T) = \fracm{I\tsub{0}}{1+R\cdot e^{\left(-\fracm{E_{A}}{k_{B}T}\right)}}
\end{align}
where, $I(T)$ is the intensity of the SPE at temperature $T$, $I\tsub{0}$ is the maximum intensity of the SPE, $k\tsub{B}$ is the Boltzmann constant, $R$ corresponds to the ratio of the radiative lifetime (\taum{}\tsub{r}) to the non-radiative lifetime (\taum{}\tsub{nr}) of the SPE, and $E\tsub{A}$ corresponds to the activation (or ionization) energy of the defect state\cite{heCascaded2016,partoDefect2021,hePhonon2016,luoSingle2019}. The energy $E\tsub{A}$ has been interpreted as the energy required to delocalize the defect-bound exciton into the free exciton band\cite{partoDefect2021,luoSingle2019,stevensEnhancing2022,wuDefect2017} or as the energy required for the defect-bound exciton to transition to another nearby defect state\cite{hePhonon2016,heCascaded2016}. The equation provides three possible pathways to improve the working temperature of the SPEs: reducing \taum{}\tsub{r} - the radiative lifetime of the SPE, increasing \taum{}\tsub{nr} - the non-radiative lifetime of the SPE, and increasing $E\tsub{A}$ - the confinement potential. \textcite{luoSingle2019} used a combination of plasmonic nanocavities (to decrease \taum{}\tsub{r}) and flux-grown high-quality WSe\tsub{2} crystal with lower defect density (thus increasing \taum{}\tsub{nr}) to demonstrate single photon nature (\gtzero{} - 0.35) up to 160 K for their emitter. Similarly, \textcite{partoDefect2021} used hBN-capped high-quality WSe\tsub{2} monolayers with low intrinsic defect density (increasing \taum{}\tsub{r}) and employed an electron beam to create deeper defect states within the band gap (increasing $E\tsub{A}$) to demonstrate the single-photon nature (\gtzero{} - 0.27) of their emitter up to 150 K. \textcite{stevensEnhancing2022} used gate voltage to suppress the background emission, which helped them achieve high signal-to-noise ratio (SNR) even when SPE intensity dropped with temperature, facilitating a demonstration of single-photon nature (\gtzero{} \simm{}0.47) up to 110 K. \textcite{gavinHighTemperature} performed covalent diazonium functionalization of graphite in layered WSe2/graphite heterostructures to quench the emission from the shallow defect states while maintaining the high emission strength of the SPE. With this technique, they demonstrated the first SPE implementation with $>$ 90\% purity (\gtzero{} $<$ 0.1) above 10 K temperature. They were also able to demonstrate single photon nature (\gtzero{} - 0.21) up to 110 K for their emitter.\\

\textcite{hePhonon2016} claimed that phonon-mediated processes being the explanation behind the temperature dependence of both the linewidth and the intensity, is supported by the fact that they observed similar values of $E\tsub{LO}$ and $E\tsub{A}$ for one of their emitters. But this observation is not universal, and the values of $E\tsub{LO}$ and $E\tsub{A}$ have been measured to be quite different from each other\cite{heCascaded2016,partoDefect2021}.

% --------------------------------------------------------------
\section*{Trends in reported TMDC-based SPE figures of merit}{\label{Trends}}

In the following, we  plot data from multiple SPE implementations using TMDCs\cite{palacios-berraqueroAtomically2016,sortinoBright2021,pengCreation2020,partoDefect2021,luoDeterministic2018,brannyDeterministic2017,mukherjeeElectric2020,wangHighly2021,peyskensIntegration2019,palacios-berraqueroLargescale2017,wuLocally2019,shepardNanobubble2017,srivastavaOptically2015,soPolarization2021,caiRadiative2018,koperskiSingle2015,heSingle2015,tonndorfSinglephoton2015,tripathiSpontaneous2018,kumarStrainInduced2015,iffStrainTunable2019,chhaperwalSimultaneously2024,cianciSpatially2023,abramovPhotoluminescence2023,blauthCoupling2018,caiChargedepletionenhanced2024,caiCoupling2017,Drawer2023-qf,errando-herranzResonance2021,flattenMicrocavity2018,gaoAtomicallythin2023,heCascaded2016,vonhelversenTemperature2023,kimPosition2019,kumarResonant2016,paurResonant2020,ripinTunable2023,utamaChemomechanical2023}. \textcolor{\mcolor}{A table summarizing the measured SPE parameters from these implementations is given in \ref{Suppl:SPEparams}.} Due to a lack of information about the specific measurement conditions and the data processing techniques used \textcolor{\mcolor}{(see \ref{Suppl:SPEparams})}, an exact comparison among these works may be difficult. But the overall trends emerging from these plots are still useful in understanding the development and the underlying physics.
\subsection*{Brightness of SPEs over the years}{\label{cpsVSyear}}
\autoref{fig:trends}a shows the trend of SPE implementations getting progressively better over the years in terms of their emission rates. Cavity-based implementations are observed to be superior in this regard as they directly improve radiative rates of the emitters.
% ------------
\subsection*{Brightness versus purity}{\label{cpsVSg20}}

\autoref{fig:trends}b shows the emission rates of various SPEs against their respective single photon purities. We can infer two trends from the plot:\\
1. Overall, \gtzero{} value decreases as the emission rate is increased. The purity of the SPE depends on the ratio of the single photon emission rate to the total detection events happening on the SPADs. These detection events include the dark counts of the SPADs, ambient light leaking into the interferometer, and emission from other emitters in the sample. An increased SPE emission rate improves this ratio and, thus, the purity measured in the \gttau{} measurement. Thus, the \gtzero{} value is expected to be lower for SPEs with a higher emission rate. Note that this trend holds true when comparing the brightness and the purity of multiple SPE implementations; however, it does not apply when comparing the purity of a particular SPE implementation at different emission rates. For a particular emitter, purity degrades with increasing emission rate as \revised{mentioned} in the \revised{following detailed discussion on \nameref{Brightness}}.\\
2. \textcolor{\mcolor}{On an average, cavity-based implementations have a degraded single photon purity [higher \gtzero{} value]} compared to non-cavity-based SPEs for similar emission rates. Photonic cavities are implemented to improve the SPE emission rates via the Purcell effect. \textcolor{\mcolor}{In general, the Q-factor of these weak-coupling cavities is modest, with a relatively broad bandwidth}. This allows multiple emitters to be coupled to the cavity if they are spectrally neighbouring. The design wavelength of the cavities may not precisely match the SPE emission wavelength due to local environmental fluctuation of defect energies at different emitter locations or fabrication accuracy limitations. These two effects together mean that while SPE's emission rate increases, its SNR (Single photon emission divided by total emission from other nearby sources) may degrade with photonic cavities.

\subsection*{Brightness versus lifetime}{\label{cpsVStau}}
\autoref{fig:trends}c shows SPE emission rates versus their respective lifetimes. We note that the maximum emission rate of the SPEs increases with a decrease in their lifetime. The emission rate of an emitter is inversely proportional to its radiative lifetime. Therefore, this trend suggests that the decrement in lifetime across multiple implementations possibly stems from the predominant decrement in their radiative lifetime. Still, the emission rates in all of these implementations are far below their lifetime-limited rates (\autoref{fig:trends}c). Thus, we conclude that a major portion of the reported lifetime in SPEs in the literature is due to their non-radiative component, which is significantly larger than the radiative component.

We also note that the lifetimes of SPEs within photonic cavities are, on average, smaller compared to those without cavities. This is expected as cavities decrease the radiative lifetimes through Purcell enhancement, reducing the overall lifetime as well. \\

% ---------------------------------
\subsection*{Lifetime versus linewidth}
\begin{figure}[!hbt]
	\centering
	\vs{-0.25in}
	\hs{-0in}
	\includegraphics[width=1.2\paperwidth,center]{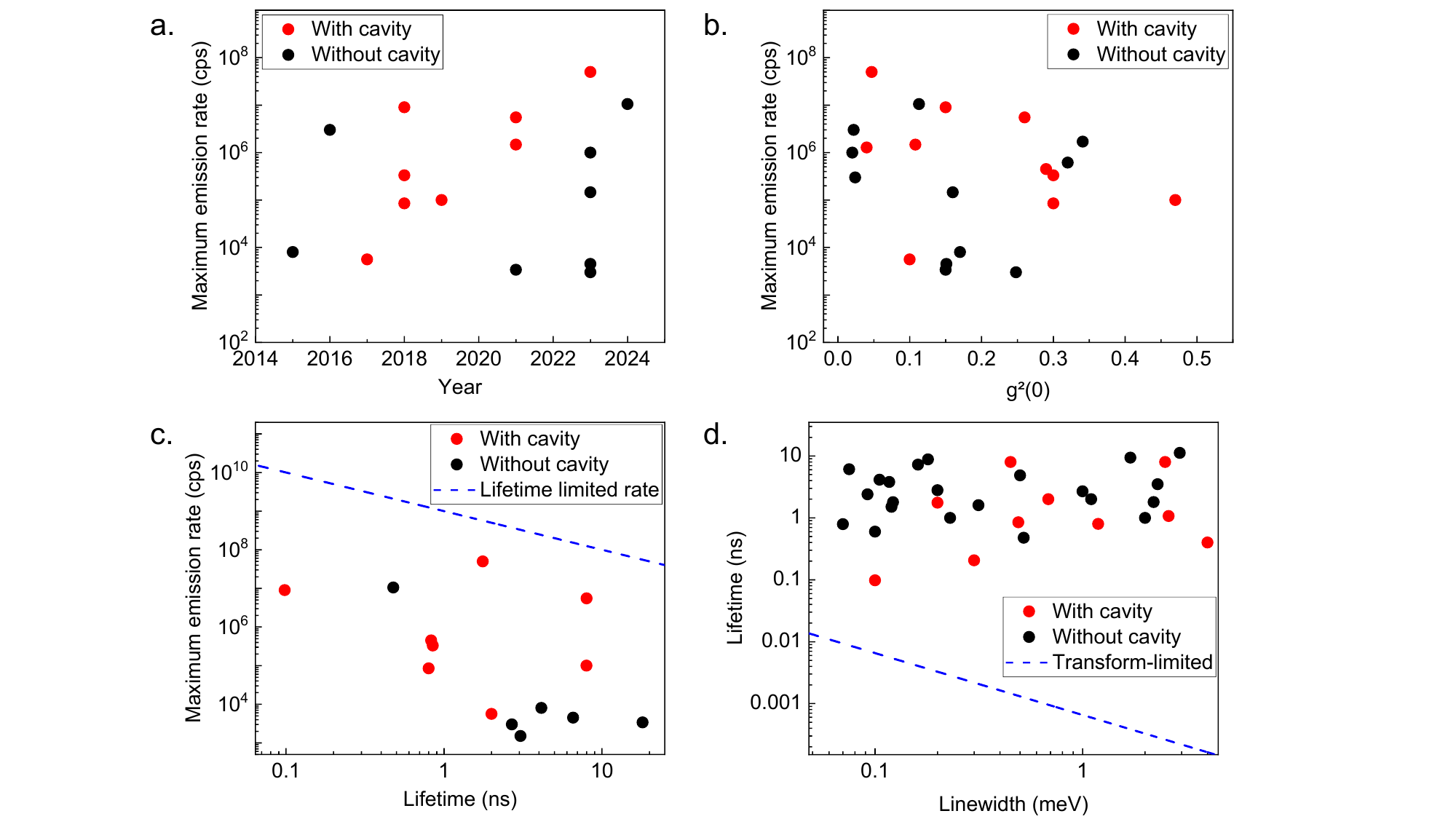}
	\vspace{-0in}
	  \caption{\textbf{Analysis of various trends in the figures of merit for different SPE implementations in literature.} (a) The maximum emission rate is plotted against the year of publication for different SPE implementations. (b) Maximum emission rate is plotted against the respective \gtzero{} values reported for various SPE implementations. (c) The maximum emission rate is plotted against the respective lifetimes reported for various SPE implementations. Blue dashed line marks the predicted emission rate if the reported lifetimes were assumed to be the radiative lifetimes of the emitters. (d) Lifetime is plotted against the respective linewidths reported for multiple SPE implementations. Blue dashed line denotes the Fourier transform limited lifetime-linewidth relation if the reported lifetimes are assumed to be radiative lifetimes of the emitters. In all the plots, data from implementations with (without) cavities are shown in red (black) circles.}\label{fig:trends}
\end{figure}

{\label{tauVSlinewidth}}
\autoref{fig:trends}d shows the lifetime of different SPEs against their respective linewidths. Ideally, in the absence of any dephasing mechanisms and external perturbations, the linewidth and lifetime of an emitter should be inversely related through the energy-time uncertainty relation\cite{sortinoBright2021,vonhelversenTemperature2023,moodyIntrinsic2015}:
\begin{align}\label{eq:lifetime_linewidth_transform}
\gammam{} = \fracm{\hslashm{}}{\taum{}}
\end{align}
where, \gammam{} is the linewidth and \taum{} is the lifetime of the emitter. But there is no discernible correlation between the lifetimes and linewidths of the SPEs in the \autoref{fig:trends}d. Also, the lifetimes of all the emitters are far greater than the ones suggested by \autoref{eq:lifetime_linewidth_transform} (\autoref{fig:trends}d). We can thus conclude that the linewidths of the SPEs are broader than the uncertainty-limited widths. This may be due to decoherence, random spectral jitter, and other interactions with the environment.
\section*{Proposed reporting procedure for figures of merit of TMDC-based SPEs}{\label{Charac_procedure}}
In this section, we first discuss the figures of merit of SPEs for their use in scalable quantum technologies. We also propose characterization and reporting procedures to facilitate the objective comparison of different SPE implementations, as well as to evaluate their performance as building blocks for various quantum technologies. Next, we plot these figures of merit from the literature across multiple SPE implementations, shedding light on some of the trends in these plots. Such trends may be beneficial for determining the path forward towards making TMDC-based SPEs more suitable for quantum technologies. \\

% ---------------------------
\subsection*{Brightness}{\label{Brightness}}
The brightness of an SPE is a crucial figure, as it directly dictates the operation speed of the systems utilizing the SPEs. Brightness should ideally be as high as possible and limited only by the radiative lifetime of the emitter. Within literature, brightness of an SPE is often defined in two ways:\\
\textbf{Quantum efficiency and excitation laser pulse rate:} If the excitation source for the SPE is a pulsed laser, the fraction of pulses resulting in a photon emission can be described as its quantum efficiency (QE).  This is often reported as a percentage\cite{sortinoBright2021,caiChargedepletionenhanced2024}. Ideally, exactly one photon should be emitted by the excitation of each laser pulse, even when the pulse repetition rate is high. QE directly corresponds to the internal quantum efficiency (IQE) of the defect state functioning as SPE when the collection efficiency is 100\%. IQE of the defect state is the ratio of its radiative decay rate to its total decay rate. When photon outcoupling is perfect, the IQE of the SPE is the limiting factor for brightness. However, when the collection efficiency is not perfect, the photon collection loss contributes to the total loss of photons, in addition to the non-radiative component of the lifetime. In such cases, the QE measured as the fraction of photons collected versus pulses fired does not correspond to the IQE and is less than the IQE. With an increase in the excitation pulse repetition rate, the brightness is eventually limited by the radiative lifetime of the SPE, even with perfect outcoupling and IQE. Quoting high quantum efficiency at high repetition rate is desirable.\\
\textbf{Photon emission rate:} The maximum number of photons emitted from the SPE can be taken as its brightness in counts per second. The emission rate from the SPE increases linearly with lower powers initially and saturates at higher powers. The saturation value is the maximum emission rate of the SPE. This can be measured using either a pulsed or a continuous-wave (CW) laser.
\\
Either of these can be experimentally determined in two ways: SPAD counts and integrated PL counts.\\
\textbf{SPAD counts:} The emission from the SPE is spectrally filtered around the SPE energy through a band pass filter and incident on an SPAD \cite{panglosse_modeling_2021, georgieva_absolute_2021}. The click rate of the SPAD, adjusted for the losses in the measurement setup, gives the brightness of the SPE.\\
\textbf{Integrated PL counts:} The PL spectrum can be integrated around the SPE peak of interest and adjusted for accumulation time and other losses in the system to give the brightness of the SPE.

While exciting the SPE, the whole device area under the diffraction-limited laser spot also gets excited and emits photons. Most of it can be spectrally filtered out, but some emitters (which need not be quantum emitters) may lie close to the SPE in energy, and the tails of their emission may come in the same energy window as the SPE. SPEs show a strong saturation with power, but these other emitters may not. This causes the SNR to degrade with increasing excitation laser power (see supplementary information note 3 of Ref. \cite{chhaperwalSimultaneously2024}). This may result in an erroneous overestimation of the brightness of the SPE if special care is not taken to remove the contribution of other emitters from the measured data. This may not be possible using the method that employs a spectral filter and SPAD to measure brightness. The reason is that the tail of other emitters will have energetically similar photons as those from the SPE and thus cannot be filtered out by the band-pass filter. The solution is to use the second method, i.e., integrated PL counts as follows: The PL spectra at a particular power can be fitted with a baseline and multiple Voigt functions corresponding to each peak. The main SPE peak can thus be taken, which has contributions from other peaks and background removed. This peak can then be integrated within its FWHM to give the pure SPE emission rate (see Figure 3c of Ref. \cite{chhaperwalSimultaneously2024}). This method can be used irrespective of the excitation scheme (pulsed or CW). In cases where SPAD count rate is presented as the SPE emission rate, reporting of \gtzero{} value at or close to this emission rate is desirable.

For calculating the true collected photon rate of the SPE from the above-mentioned experimental value, the efficiency of the measurement setup is crucial. However, this efficiency should not include the collection efficiency of the setup, which is an important metric in its own right (discussed later).

\subsection*{Purity}{\label{Purity}}
The \gtzero{} value should be as low as possible, ideally approaching zero. As discussed in the last section, the SNR of an SPE degrades with excitation power (and thus, its emission rate). Therefore, a high emission rate of an SPE does not have a practical benefit if it does not maintain its single-photon purity at that rate. The measurements of \gtzero{} at low emission rates are thus not very useful when comparing across multiple implementations. Thus, the \gttau{} measurement should ideally be performed at the maximum emission rate or at least close to it. This emission rate should also be mentioned along with the \gtzero{} value. \gttau{} measured at multiple powers gives an even better understanding of the robustness of the SPE purity at higher emission rates.

The timing jitter of the SPADs used in the measurement can influence the measured \gtzero{} value. A setup with SPADs having higher timing jitter can erroneously report coincidences, degrading the measured \gtzero{} value \cite{hadfield_single-photon_2009, gong_jitter-calibrated_2022}. Therefore, to get an estimate of the SPE's objective purity, it is advised to deconvolute the \gttau{} curve from the Gaussian instrument response function (IRF) of the setup. This can be done by fitting the measured \gttau{} with a convolution of two functions: the ideal \gttau{} equation given by $1-ae^{-t/\taum{}}$ and a Gaussian function given by $\frac{1}{\sigma \sqrt{2\pi}} e^{\frac{-t^2}{2 \sigma^2}}$ where \sigmam{} is the overall timing jitter (root mean square of individual component timing jitter values)\cite{chhaperwalSimultaneously2024,tripathiSpontaneous2018,heCascaded2016}. This becomes increasingly important when the lifetime of the SPE is comparable to or lower than the detector’s timing jitter.

If the dark count rate of the SPADs is comparable to the detection rate of incoming photons, erroneous coincidences artificially degrade the \gtzero{} value \cite{mcguinness_quantum_2010,cova_avalanche_1996}. Therefore, SPADs with lower dark count rates relative to the SPE emission rates facilitate the measurement of the intrinsic single-photon purity of the source. The efficiency of the SPADs at the wavelength of the emitter determines the accumulation time required for the \gttau{} measurement. Since the setup contains two SPADs, the required measurement duration grows as the inverse square of the SPAD efficiency for the same emitter. Thus, SPADs with higher efficiency at the wavelength of the SPE are desirable for time-constrained measurements. Other components, such as the bandwidth of the bandpass filters used for \gttau{} measurement, can also influence the measured value of \gtzero{}. A filter with a broader pass band will allow more photons from the energetically neighbouring peaks to pass through and degrade the measured \gtzero{} value. Thus, technical details of all the components used in measurement should be specified for a clear and objective assessment of different implementations of SPEs.

\subsection*{Indistinguishability}{\label{Indistinguishability}}
For many applications, such as interferometry-based quantum computing, single photons must be indistinguishable from one another. The indistinguishability of photons from an SPE can be quantified by performing a Hong–Ou–Mandel (HOM) measurement \cite{hongMeasurement1987} in a Mach-Zehnder interferometer. This measurement may not always be feasible. It is known that a high degree of polarization, low spectral jitter, and low timing jitter are required to achieve indistinguishability. Therefore, the degree of polarization (DOP), along with the lifetime and linewidth of the SPE, can be used to gauge the indistinguishability of the SPE. We propose that polarization-resolved PL measurements be carried out and the calculated DOP be reported for all SPE implementations\cite{chhaperwalSimultaneously2024,heCascaded2016,partoDefect2021,wangHighly2021,dangIdentifying2020,linhartLocalized2019,xiangMagnetic2025,seratidebritoProbing2024,yangRevealing2023,clarkSingle2016,heSingle2015,debritoStrain2022,luOptical2019}. Again, the SPE intensity used in the DOP calculation should be the SPE's integrated counts from PL after removing the background and contribution from energetically neighbouring peaks.

Another important characteristic of the SPE is its coherence time, which is directly related to its indistinguishability. The visibility of the fringe contrast in the HOM measurement can be estimated as $T\tsub{2}/2T\tsub{1}$ where $T\tsub{2}$ is the coherence time and $T\tsub{1}$ is the lifetime of the emitter\cite{vonhelversenTemperature2023}.
\section*{A survey of SPEs in quantum technology applications and identifying research needs for TMDC-based SPEs}{\label{SPEs_in_QTech}}

SPEs offer the foundational hardware for different photonic quantum information processing applications. As single photons are discrete quanta, they are easy to be involved in encoding, manipulation, and computing. Recent studies have used single photons’ indistinguishability and long coherence time to develop quantum technologies, including computation \cite{maring_versatile_2024}, communication \cite{gaoAtomicallythin2023}, simulation \cite{wang_deterministic_2023}, metrology \cite{motes_linear_2015, von_helversen_quantum_2019}, and secure information processing \cite{couteau_applications_2023}. Experimentally, $P$, indistinguishability ($V_{\mathrm{HOM}}$), and $B$ of single photons contribute to the success of its applications, where different applications ask for different weights of these parameters. Applications focusing on multi-photon interference are limited by the $V_{\mathrm{HOM}}$ and the SPE’s spectral stability, where cryptographic tasks and information processing require higher purity, coherence, and rate of generation. We here discuss how such applications incorporate the established deterministic technologies of SPEs and where TMDCs stand as a technology for achieving experimental validation and industrial deployment. \\

% -------------------------------------------
\subsection*{Linear optics quantum computing}
Implementation of single photons along with some linear optical components (beam splitters, phase shifters) has been widely demonstrated to compute quantum logics \cite{Knill2001}. This process is popularly known as Linear-optics quantum computing (LOQC). It requires a stream of indistinguishable single photons, which interfere quantum mechanically in multi-port optical networks \cite{wayo_linear_2025}. Early demonstrations of LOQC relied on probabilistic spontaneous parametric down-conversion (SPDC) sources, as reviewed by Franson \etal{}\cite{franson_experimental_2003}. However, recent developments have shifted towards a more scalable approach using single photon sources, such as QDs.

For an SPE to serve a better candidacy for LOQC, it requires simultaneously having high $P$, high $ V_{\mathrm{HOM}}$ interference visibility between consecutive photons or photons from different sources,  and high $B$. Semiconducting QDs like epitaxially grown III-V compound QDs are considered as good platforms for LOQC, with demonstrations reporting $P > 0.99$ and $V_{\mathrm{HOM}} > 0.9$ \cite{Somaschi2016, ding_-demand_2016}. Nanophotonic waveguides with embedded QDs have been demonstrated to generate high-purity single photons with excellent coupling efficiency \cite{kirsanske_indistinguishable_2017}. Loredo \etal{}  realized an electrically controlled QD-micropillar based SPE which achieved high purity (97–99\%) and high indistinguishability (up to 96\% under resonant excitation), and absolute brightness of 14\% \cite{loredo_scalable_2016}. They also indicated towards the prospect of implementing LOQC with their SPE. In 2025, Ding \etal{}  observed that InAs QDs, when coupled to a microcavity, can be used as a single photon source that surpasses the loss-tolerant threshold for scalable LOQC \cite{ding_high-efficiency_2025}. It also achieved a system efficiency of 0.712, alongside \gtzero{} = 0.0205 and photon indistinguishability of 0.986.  In a different study, Maring \etal{} from the \emph{Quandela} reported a cloud-accessible photonic quantum computing platform based on an on-demand InAs QD source that feeds a reconfigurable linear-optical processor \revised{(\autoref{fig:qt_tech1}a)}. Their system demonstrated single-photon purity over 99\%, pairwise $ V_{\mathrm{HOM}}$ of 91–94\% across all photon pairs. \cite{maring_versatile_2024}. Researchers in 2025 have also explored using a single spin–photon interface to emit photons sequentially into graph/cluster states using QD SPEs to run computations on these states rather than requiring two-photon gates \cite{meng_temporal_2025, huet_deterministic_2025}. However, fabrication and experiments require cryogenic temperatures and precise growth control, which often hinders scalability. 

 Recent literature report SPEs with TMDCs can routinely achieve $g^{(2)}(0) \simm{} 0.02-0.1$, which indicates high single photon purity comparable to QDs \cite{ paralikisTailoring2024, palacios-berraqueroLargescale2017, chhaperwalSimultaneously2024} with promising brightness values. However, from the perspective of LOQC, TMDC-based SPEs still lack the required photon indistinguishability due to strong exciton–phonon coupling and spectral diffusion. While cavity coupling and electrical control are being explored to address these limitations, reproducible demonstrations of near-unity $ V_{\mathrm{HOM}}$ are yet to be reported.
 
 %------------------------------
\subsection*{Boson sampling}
Boson sampling, first suggested by Aaronson and Arkhipov, is a sampling from the output distribution of \revised{$N$}  single photons (ideal) propagating through an \revised{$M$}-mode linear interferometer (\revised{$M \gg N$}) \revised{(\autoref{fig:qt_tech1}b)} \cite{Aaronson2013}. Its computational hardness originates from many-photon interference, which is sensitive to photon distinguishability and loss. While classically it is very hard to simulate, a dedicated photonic hardware comprising SPEs, linear optical elements, and single-photon detectors can mitigate this issue. Thus, boson sampling indicates quantum computational advantage without requiring a universal quantum computing architecture \cite{Brod2019}. For SPEs, boson sampling also requires a similar platform, as it did for \revised{LOQC}. First, the emitter should emit highly pure single-photon states; multi-photon states can introduce background noise, while also altering the effective input state and biasing the sampler. Second, photons must be highly indistinguishable; this tracks how much multi-photon interference survives. Third, the photons must be delivered into the interferometer with high end-to-end efficiency (often first-lens or fiber-coupled brightness). Finally, the emission must be spectrally stable and mode-matched. Otherwise, $ V_{\mathrm{HOM}}$ collapses over the integration time \cite{rakhlin_demultiplexed_2023, Gard2015}. 

Experimentally, so far, the strongest SPE candidates for boson sampling have been semiconductor QDs, as such platforms, in resonant excitation with cavity coupling, have shown to deliver high $P$, high $ V_{\mathrm{HOM}}$, and usable brightness simultaneously \cite{maring_versatile_2024}. In a 2017 study, Wang \etal{} showed high-efficiency multi-photon boson sampling using a resonantly driven InAs/GaAs QD micropillar. They also incorporated active temporal-to-spatial demultiplexing to increase the efficiency. The source simultaneously achieved \gtzero{}=0.027, and a $ V_{\mathrm{HOM}}$ of 0.939 for photons separated by 13 ns and 0.900 even for separations as high as 15 $\mu$s \cite{Wang2017}. Another related work by Loredo \etal{} realized boson sampling with true single photon Fock states using a demultiplexed quantum-dot source with  $P=0.990\pm0.001$ \cite{loredo_boson_2017}. A milestone in these developments is the 20-input-photon experiment in a 60-mode interferometer, which can detect up to 14 photons at the output, which are then sampled over a Hilbert space of size $\sim{10}^{14}$ \cite{Wang2019}. The core of this study is based on single photon streams produced from an InAs/GaAs QD  deterministically coupled to a micropillar cavity. Several other studies have shown promise while incorporating time bin encoding or deliberate photon loss while performing boson sampling using QD sources \cite{He2017, Wang2018d}.

Similar to the other two platforms, the lack of demonstration of high $V_{\mathrm{HOM}}$ while maintaining high $P$ and $B$ and low spectral jitter remains a critical bottleneck for TMDC-based SPEs to be used in a Boson sampling architecture. Although some recent studies on TMDCs have shown some observable two-photon interference, they are still incapable of operating in the high $ V_{\mathrm{HOM}}$ regime ($\sim 1$) required for large $-n$ boson sampling \cite{wyborski_toward_2025}.

%--------------------------------
\subsection*{Quantum simulation and quantum walks}
The requirements for efficient quantum simulation or quantum walks implemented with SPEs are similar to LOQC \revised{and boson sampling}. Quantum walks using single photons primarily demand purity and brightness, while multi-photon walks and simulations are governed by indistinguishability and loss. \revised{\autoref{fig:qt_tech1}c is a simplistic depiction of a two-photon quantum walk, only using beam splitters.}

Historically, quantum walks have been demonstrated mostly using heralded photons \cite{peruzzo_quantum_2010, tang_experimental_2018, broome_discrete_2010}. Recent developments indicate that quantum dots can serve as a scalable alternative for this application, as they can deliver high purity, high $ V_{\mathrm{HOM}}$, and chip-compatible integration simultaneously. A study by Wang \etal{} observes multi-photon quantum simulations, such as bosonic suppression and entanglement generation, when a QD single-photon source with a low-loss SiN programmable PIC is directly aligned with quantum walk unitaries as programmable interferometers \cite{wang_deterministic_2023}. Another study emphasizes that SPEs may no longer be limited to single-photon benchmarks, as they can generate on-demand entanglement seeds with high indistinguishability \cite{meng_deterministic_2024}. One example of such entangled states is a three-qubit (spin + two photons) entanglement. This study also reports single-photon indistinguishability exceeding 90\% and directly verifies genuine three-qubit entanglement without background subtraction via fusion operations, making it a potential choice for quantum walks. Moreover, QDs, primarily under resonant excitation, offer deterministic photon cluster state creating capability with single photon $g^{(2)}(0) < 0.01$ and $ V_{\mathrm{HOM}} > 0.9$, which makes them promising for continuous time quantum walk (CTQW) \cite{istrati_sequential_2020, hauser_deterministic_2025}. Researchers also explore other techniques, such as passive demultiplexing, nowadays to extract multi-photon states from QD-based SPE generated single photon stream for quantum walks \cite{karli_passive_2025}.

Considering the benchmarks, TMDC-based SPEs are strong candidates for single-photon quantum walks due to their high brightness and purity, but are still catching up for multi-photon quantum walks/simulations (which require high indistinguishability). A recent study reports two-photon interference (HOM) measurements in a triggered architecture where Wyborski et al. observed MoTe$_2$ as a strain/defect-engineered emitter in the near-IR (1090–1200 nm)  \cite{wyborski_toward_2025}. With a reported $P>0.9$ and a $ V_{\mathrm{HOM}} \sim 10\%$, potentially reaching $\sim 40\%$ with temporal filtering, this study shows the hope of achieving higher and potentially required $V_{\mathrm{HOM}}$ for multi-photon quantum walk using TMDCs.

%---------------------------------
\subsection*{Quantum teleportation}
Quantum teleportation using photonic architecture is a qubit (information) transfer scheme using shared entanglement between two distant nodes (Alice and Bob) and a Bell-state measurement (BSM) at Bob's end \cite{bennett_teleporting_1993}. The communication is secured by two-photon interference \cite{lo_unconditional_1999}. The teleportation fidelity is therefore governed by $P$, $ V_{\mathrm{HOM}}$, and the fidelity of the on-demand entangled photon pair when SPEs are used. Experimentally, quantum teleportation is realized when an average fidelity exceeds the classical limit of 2/3, while the security and the usable communication rate are limited by the probabilistic BSM and source brightness \cite{massar_optimal_1995, bennett_teleporting_1993}.

The first report of experimental quantum teleportation using an entangled photon pair came in 1997 when Bouwmeester \etal{} demonstrated transmission and reconstruction of entangled polarization states of single photons from non-linear probabilistic photon generation \cite{bouwmeester_experimental_1997}. But the first significant progress towards using SPE for quantum teleportation came in 2003 by Fattal \etal{} \cite{fattal_quantum_2004}. The study employed a triggered InAs QD emitter to achieve the teleportation of dual-rail photonic qubits, although its performance was limited by the reduced $ V_{\mathrm{HOM}}$ and collection efficiency. Building on this finding, in 2018, Reindl \etal{} demonstrated all-photonic quantum teleportation using GaAs QDs as sources of both single photons and entangled pairs \cite{reindl_all-photonic_2018}. 

The most recent developments in photonic quantum teleportation focus on the feasibility of networking and utilizing dissimilar sources for the same purpose. A 2020 report by Anderson \etal{} presented telecom C-band InAs/InP QDs as the source for single and entangled photons that can help realize photonic quantum teleportation of polarization-encoded qubits \cite{anderson_quantum_2020}. The study further reported an average post-selection fidelity of up to $88.4\pm4.0\%$, while achieving long single photon coherence times exceeding 1 ns. Using a resonantly driven QD (Quandela \emph{e-Delight}) SPE, Polacchi \etal{} demonstrated photon-number-encoded teleportation of a single-rail vacuum–one-photon qubit \revised{(\autoref{fig:qt_tech1}d)} \cite{polacchi_teleportation_2024}. A November 2025 report by Strobel \etal{} demonstrates all-photonic teleportation using frequency converted (to telecom) photons with conjugate polarizations from two distant NIR (780 nm) GaAs QDs \cite{strobel_telecom-wavelength_2025}. They report a realized teleportation fidelity of 0.721 while implementing resonant excitation of the biexciton-exciton radiative cascade. A 2025 paper reports quantum teleportation with a fidelity of 0.82, employing two dissimilar GaAs QDs to address the fabrication issue and feasibility \cite{laneve_quantum_2025}. 

As stated in the earlier sections, TMDCs offer a promising platform for SPEs. However, as of now, there are very few experimental demonstrations of quantum teleportation using TMDC-based SPEs. Recent advances have shown deterministic polarization control of the photons generated from such SPEs, paving the way for realizing TMDC-based emitters in teleportation. However, their present coherence and indistinguishability metrics currently limit immediate teleportation-grade performance \cite{paralikisTailoring2024, paralikisTunable2025}. 

%-----------------------------------------------
\subsection*{Quantum key distribution: BB84, E91}
Quantum key distribution (QKD) is a discrete-variable quantum communication scheme where two remote parties (named as Alice and Bob) set a shared secret key that is secured under quantum measurement disturbance and nonclassical correlations. The two main protocols are --- BB84 (prepare and measure) and E91 (entanglement based). These were introduced in the 1980s and 1990s and remain the reference points for modern QKD \cite{bennett_quantum_2014, ekert_quantum_1991}. Often, practical deployments utilize weak coherent pulses (WCPs) with decoy states to conveniently manage security. 

In the BB84 protocol, for example, Alice sends one of four polarization encoded qubit states: \[{|H\rangle,\ |V\rangle,\ |D\rangle=\tfrac{|H\rangle+|V\rangle}{\sqrt2},\ |A\rangle=\tfrac{|H\rangle-|V\rangle}{\sqrt2}}.\]
Bob then measures on a randomly chosen basis either $HV$ or $DA$. Most importantly, BB84 assumes one photon per signal for its security; otherwise, multi-photon events attract photon number splitting attacks \cite{shor_simple_2000, tupkary_qkd_2025}. In the E91 protocol, proposed by Ekert in 1991, a source prepares entangled photon pairs, ideally in a Bell state like
\[|\Phi^+\rangle = \tfrac{|H\rangle_A|H\rangle_B + |V\rangle_A|V\rangle_B}{\sqrt2}.\]

Alice and Bob then choose a measurement setting, from which a subset generates the raw key, and another subset is used to test the Bell inequality. It requires high-quality entanglement distribution with sufficient pair generation rate and low noise, which often creates a practical burden \cite{ekert_quantum_1991}. 

QKD prioritizes multi-photon suppression, channel brightness, and wavelength with low-loss links over indistinguishability \cite{lo_measurement-device-independent_2012}. QKD can tolerate modest nonzero \gtzero{} values, but it is rate-limited by brightness. Fiber QKD prefers the telecom bands (O/C/L), while free-space links can tolerate visible/NIR. As of the latest developments, many SPEs operate outside the telecom band. Thus, they require quantum frequency conversion that preserves the encoding fidelity and doesn't inject excess noise. QKD also requires stable encoding and the least spectral diffusion to reduce the quantum bit error rate (QBER). 

\begin{figure}[!hbt]
	\centering
    \includegraphics[width=0.9\paperwidth,center]{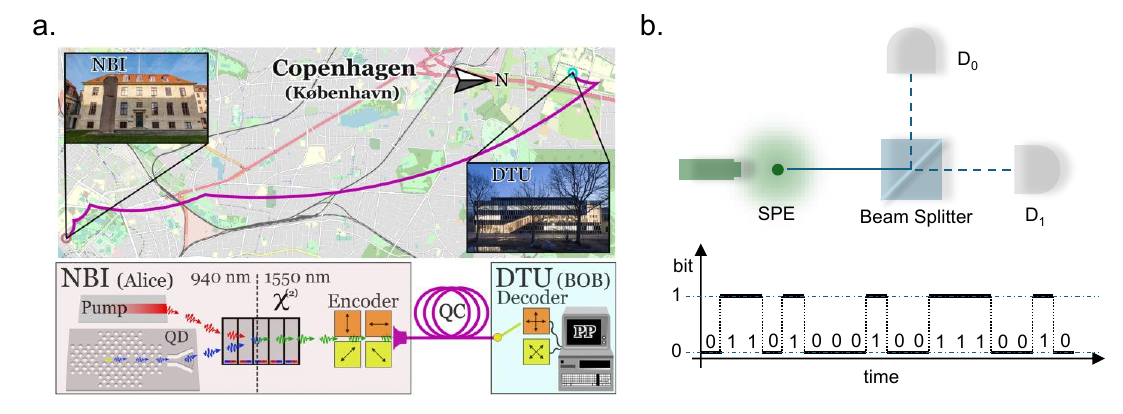}
    \captionof{figure}{\revised{\textbf{An overview of QKD and QRNG using single photon emitters.} (a) Schematic of a QKD experiment. The map shows the quantum channel spanning 18.1 km. A schematic of the experiment using optical components is also shown. Reproduced with permission from ref. \cite{zahidy_quantum_2024}. Copyright 2024, under CC BY 4.0 license. (b) A simple implementation of a quantum random number generator by using one SPE, one beam splitter, and two single photon detectors ($D_0$ and $D_1$).}}
    \label{fig:qt_tech2}
\end{figure}

QKD with single photons exploiting QD-micropillar cavity SPE was first documented in 2009, with a maximum secure key rate of
160 bit/s and a measured QBER of 5.9\% \cite{intallura_quantum_2009}. Takemoto \etal{} demonstrated the first telecom-C band (1560 nm) QKD using InAs/InP QDs \cite{takemoto_transmission_2010}, and later extended the QKD operational range to 120 km using superconducting nanowire single photodetectors in 2015 \cite{takemoto_quantum_2015}. In 2023, a 175 km demonstration of QKD using frequency-converted (to telecom 1550 nm) 980 nm-emitting InGaAs/GaAs QD SPE was reported \cite{morrison_single-emitter_2023}. In 2024, an 18 km dark fiber BB84 field trial, \revised{stretching from the Niels Bohr Institute (NBI) to the Technical University of Denmark (DTU), by Zahidy \etal{} used InAs QD embedded in GaAs membrane (\autoref{fig:qt_tech2}a) \cite{zahidy_quantum_2024}. They reported a secret key rate of over 2 kbit/s.} Yang \etal{} performed a 79 km intercity BB84 experiment using a 1555.9 nm InAs/InGaAs/GaAs QD emitter integrated with a circular Bragg grating \cite{yang_high-rate_2024}, reporting a very low QBER (~0.65\%). Barnes \etal{} recently reported 227 km long QKD incorporated with decoy-state techniques \cite{barnes_decoy-state_2025}. In 2025, a report further pushed the “why single photons for QKD?” argument by showing that SPE QKD can surpass a weak coherent state secure-key-rate limit in a realistic channel \cite{zhang_experimental_2025}, achieving a secure key rate of $1.08\times10^{-3}$ bits per pulse over a $\sim$ 14.6 dB-loss free-space channel.

Beyond QDs, there is significant effort and activity in room-temperature telecom SPE based on defects, such as, BB84 demonstration \cite{zhang_polarization-encoded_2025} and $>$ 30 km fiber telecom QKD demonstration \cite{zhang_metropolitan_2025} using GaN defect center. The latter operates in the O-band (~1305 nm) while having \gtzero{}$=0.065 \pm0.061$, and a short excited-state lifetime of $\sim$ 315 ps. A 2022 study shows the excellent potential of using a single-molecule SPE (DBT molecules in anthracene nanocrystals) at room temperature for BB84 QKD \cite{Murtaza2023}. It further reports \gtzero{} = $0.02 \pm0.01$ and a competitive secret key rate of $\sim$ 0.5 Mbps.

For E91 protocol, some early studies in 2006 showed promise for entanglement distribution using QD-based SPEs \cite{akopian_entangled_2006, stevenson_semiconductor_2006}. In 2010, Dousse \etal{}, using self-assembled InAs QDs in microcavities, achieved an entangled photon pair generation with high extraction/brightness while maintaining entanglement quality \cite{dousse_ultrabright_2010}. Jayakumar \etal{} reported demonstration of time-bin entangled photon pairs using the biexciton–exciton cascade in III/V self-assembled QDs \cite{jayakumar_time-bin_2014}. Demonstration of entanglement distribution using a telecom band SPE based on InAs/GaA QD was reported by Xiang \etal{} \cite{xiang_long-term_2019, xiang_tuneable_2020}. Basset \etal{} demonstrated an Ekert-style (modified E91) QKD protocol using a QD entangled-photon SPE (GaAs QD embedded in Al$_{0.4}$Ga$_{0.6}$As) in a field-study environment \cite{basso_basset_quantum_2021}.  In the same year, fiber-based entanglement QKD using a blinking-free GaAs QD entangled-pair source was achieved by Schimpf \etal{} \cite{schimpf_entanglement-based_2021}. They also attained an average key rate of $\sim$55 bits/s and QBER of $\sim$8.4\% while maintaining their SPE at a considerably high temperature of 20 K. An experimental milestone was set by Zhang \etal{} in 2022 after they demonstrated device-independent QKD between two distant users based on event-ready entanglement between trapped rubidium ($^{87}$Rb) atoms separated by $\sim$400 m \cite{zhang_device-independent_2022}. They further reported entanglement fidelity $\geq0.892(23)$, Bell violation $2.578(75)$, and a QBER of 7.8\%. Recently, in 2025, Langer \etal{} achieved near-maximally entangled photon pairs based on the biexciton–exciton cascade of a single GaAs QD embedded in monolithic microlenses \cite{langer_ultra-compact_2025}.

Over the last decade, researchers have also explored spin-photon entanglement for QKD. In 2012, a study reported observation of entanglement between a semiconductor quantum dot spin and a single propagating optical photon \cite{gao_observation_2012}. Laccotripes \etal{} report spin–photon (telecom C-band) entanglement using a single InAs/InP QD, achieving an entanglement fidelity of $80.1\pm2.9\%$, beating the classical limit by more than ten standard deviations \cite{laccotripes_spin-photon_2024}. Uysal \etal{} reports the demonstration of spin–photon entanglement in the telecom band (1532.6 nm) using a single Er$^{3+}$ ion integrated with a silicon nanophotonic cavity \cite{uysal_spin-photon_2025}, with spin–photon entanglement fidelity of 0.73(3), exceeding the classical bound, after transmission through 15.6 km of optical fiber. While spin-photon entanglement studies are still in their early stages for practical E91 or any other QKD demonstration, these studies have shown promise for telecom operations. 

Although very few, some interesting reports have shown that TMDCs can be projected as an important platform for QKD applications, with further developments still needed. For prepare-and-measure schemes like BB84, recent developments have shown TMDC-based SPEs with \gtzero{}$\lesssim0.05$, MHz-level count rates, and improving control over polarization selection rules through strain and nanophotonic engineering while also making it suitable for scalable integration \cite{tonndorfSinglephoton2015, heSingle2015, chhaperwalSimultaneously2024, palacios-berraqueroLargescale2017, soPolarization2021}. In particular, Gao \etal{} in 2023 reported a full BB84 QKD emulation using a strain-engineered WSe$_2$ monolayer, achieving click rates up to $\sim$67 kHz, \gtzero{} as low as 0.034 after temporal filtering, and QBER below 1\%, with tolerable channel losses exceeding 20 dB \cite{gaoAtomicallythin2023}. In parallel, recent research has paved the way for telecom-band operation, most notably through site-controlled MoTe$_2$ emitters spanning 1080–1550 nm \cite{zhaoSitecontrolled2021}. This addresses a long-standing wavelength barrier for fiber-based QKD with TMDCs.

In contrast, for entanglement-based QKD, such as the E91 protocol, TMDC SPEs currently lag behind QD platforms. So far, there have been very few experimental developments of on-demand, high-fidelity photon–photon or spin–photon entanglement from TMDC emitters that also satisfy Bell-inequality-based security. Coherence of the photons from TMDC-based SPEs still remains limited by spectral diffusion and exciton–phonon coupling \cite{koperskiSingle2015}. Despite excellent interferometric visibilities demonstrated with hBN- and TMDC-based single photons, demonstrations of on-demand, high-fidelity photon–photon or spin–photon entanglement from a TMDC-based SPE lack sufficient experimental validations \cite{vogl_sensitive_2021}.  

%--------------------------------------------
\subsection*{Quantum random number generation}

A quantum random number generator (QRNG) converts inherently random quantum measurement outcomes into information bits. In a typical photonic QRNG setup, \revised{as described in \autoref{fig:qt_tech2}b}, a single photon hits a 50:50 beam splitter and is detected in one of two outputs via detectors. The bit is the outcome of a projective measurement at the detectors, with probabilities ideally 50\% for both of them \cite{herrero-collantes_quantum_2017}. The randomness of the bits is quantified by the min-entropy $H_{min}=-{\log}_{2}(p_{max})$ \cite{sonmez_turan_recommendation_2018}. While SPEs are not automatically a better choice than non-linear-optics based photon emitters for QRNG, they offer device compactness, discrete detection, and cleaner physical modeling.

Unlike LOQC or boson sampling, QRNG does not demand high photon indistinguishability; instead, it requires suppressed multi-photon probability, i.e., very low \gtzero{}, high brightness, and temporal stability to reduce any possible biases  \cite{herrero-collantes_quantum_2017}. Blinking, spectral diffusion, or drift in coupling efficiency can add unwanted bias to the information bits.

A very direct implementation (SPE, beam splitter, detector) of QRNG  using single photons from a single NV center in diamond was reported by Chen \etal{} in 2019 \cite{chen_single_2019}. The study mentioned possible bias, drift after collecting more than 800 GB of data over 7 days. One year later, Luo \etal{} demonstrated a room-temperature defect center in GaN layer for QRNG. They achieved raw $\sim$1.8 MHz and unbiased $\sim$420 kHz after von Neumann extraction, significantly improving the data rate \cite{luo_quantum_2020}. A different study, published in the same year, reported the experimental realization of QRNG using an hBN SPE, which feeds an integrated photonic circuit \cite{white_quantum_2020}. This study achieved NIST SP800-22 validation for the random number streams \cite{rukhin_statistical_2010}. Some recent studies mention a closely related SPE route that skips the beam splitter. Hoese \etal{}, in 2022, demonstrated NIST SP800-22 validated randomness based on the random direction profile of photon emission, not incorporating any beam splitter \cite{hoese_single_2022}. A notable study in this context is the demonstration of metropolitan-scale QRNG using a triggered telecom-band InAs QD SPE over 20 km of deployed fiber and generating randomness at a remote user \cite{gyger_metropolitan_2022}. The study achieved NIST SP800-22 validated random numbers at 23.4 kbit/s, with \gtzero{} \simm{} 0.19-0.21 at the user and entropy bounds $H_{min}$ \simm{} 7–8 bits/byte after extraction. Most recently, a 2025 study by Meng \etal{} has projected their InAs/GaAs QDs as a potential platform for QRNG \cite{meng_contextuality-based_2025}.

Given the demonstrated high $B$ and $P$ for TMDC-based SPEs, coupled with their potential for compact chip-scale integration, they are a promising candidate for QRNG. However, current literature lacks demonstrations of TMDCs for QRNG applications. Owing to their planar geometry and rapidly improving brightness–purity trade-off \cite{chhaperwalSimultaneously2024}, TMDC-based SPEs hold the potential for high-throughput QRNG hardware.

\section*{Conclusion and outlook}\label{conclusion}
In summary, TMDC quantum emitters combine a unique suite of attributes — atomic thinness, deterministic strain-based positioning, ease of heterostructure fabrication, gate- and proximity-induced tunability, and compatibility with on-chip photonics — that make them promising for scalable integrated single-photon technologies.  The recent demonstrations of simultaneous improvements of brightness and purity, together with continued advances in encapsulation, cavity and waveguide integration, and heterostructure design, indicate a rapid convergence of TMDC emitters toward the performance regimes required for many quantum photonic applications.

Despite the remarkable progress, some key challenges remain to be overcome for TMDC-based SPEs. Measured coherence times (and thus indistinguishability) for the best TMDC-based SPEs still lag behind those of the highest-performing III–V quantum dots and group-IV color centers. Maintaining high single-photon purity at increasing emission rates is an ongoing effort. Improving all these parameters at elevated temperatures remains an important research direction for these SPEs. In addition to these, parallel efforts are needed to address additional requirements, such as (1) development of a consistent model to explain the atomic origin of SPEs, their fine structure, and their dependence on the electrostatic environment; (2) engineering better outcoupling solutions for improving the adaptation of SPEs in scalable technologies without loss of temporal determinism; (3) development of electrically triggered, bright, deterministic SPEs; (4) engineering devices with the capability to dynamically tune the properties of SPEs, such as their emission energy over a large range; (5) development of TMDC-based SPEs at longer wavelengths, specifically around 1550 nm, for improved integration with existing telecom infrastructure; (6) increasing the robustness of the SPEs against random environmental fluctuations to avoid spectral and intensity jitter.

\section*{Supporting Information}
Table summarizing the measured SPE parameters (used to generate the plots in \autoref{fig:trends}) from multiple implementations present in the literature.

\section*{Acknowledgements}
K.M. acknowledges support from National Quantum Mission, an initiative of the Department of Science and Technology (DST), Government of India, a grant from Indian Institute of Science under IoE, a grant from DRDO, and a grant from I-HUB QTF, IISER Pune.
\section*{Competing Interests}
The authors declare no competing financial or non-financial interests. 
\section*{Data Availability}
Data available on reasonable request from the corresponding author.
\printbibliography

% ----------------------------
\newpage
\newpage
\newpage
\newpage
\newpage
\newpage

\newpage

\newpage

% \begin{center}
% \includegraphics[width=\textwidth]{figures_AK/sch_section7.pdf}
% \captionof{figure}{\revised{\textbf{An overview of the quantum technologies using single photon emitters.} (a) Architecture of a six-single-photon quantum computer, \textit{Quandela Ascella}, which consists of a QD SPE at 5K, a demultiplexer, and thermal phase shifters. The outputs are detected by Superconducting Nanowire Single Photon Detectors (SNSPDs). Reproduced with permission from ref \cite{maring_versatile_2024}. Copyright 2024, under CC BY 4.0 license. (b) An $M$-mode multiport interferometer for boson sampling. Adapted with permission from ref \cite{huang_simulating_2019}. Copyright 2019, American Physical Society. (c) A schematic of a two-photon quantum walk involving only beam splitters. (d) Quantum states are sent through the designated beam splitter to generate a Bell state. Half of it interferes with Alice’s part. The other half of the Bell state is the teleported state, which goes to Bob. Reproduced with permission from ref \cite{polacchi_teleportation_2024}. Copyright 2024, under CC BY 4.0 license. (e) Schematic of a QKD experiment. The map shows the quantum channel spanning 18.1 km. A schematic of the experiment using optical components is also shown. Reproduced with permission from ref \cite{zahidy_quantum_2024}. Copyright 2024, under CC BY 4.0 license. (f) A simple implementation of a quantum random number generator by using one SPE, one beam splitter, and two detectors ($D_0$ and $D_1$).}}
% \label{fig:qt_tech}
% \end{center}

\begin{center}
\includegraphics[width=0.9\paperwidth,center]{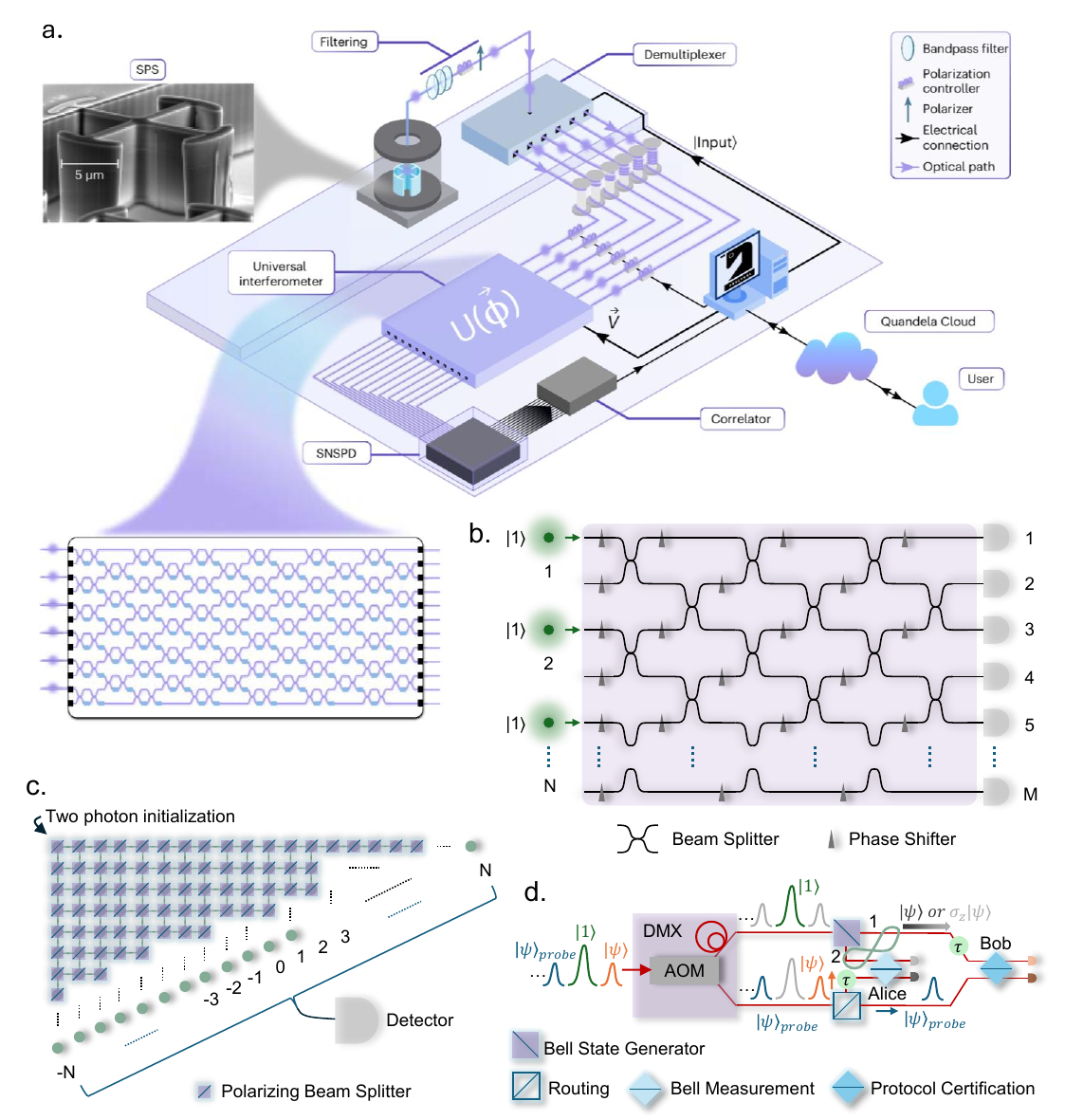}
\captionof{figure}{\revised{\textbf{An overview of the quantum technologies from LOQC to teleportation using single photon emitters.} (a) Architecture of a six-single-photon quantum computer, \textit{Quandela Ascella}, which consists of a QD SPE at 5K, a demultiplexer, and thermal phase shifters. The outputs are detected by Superconducting Nanowire Single Photon Detectors (SNSPDs). Reproduced with permission from ref. \cite{maring_versatile_2024}. Copyright 2024, under CC BY 4.0 license. (b) An $M$-mode multiport interferometer for boson sampling. Adapted with permission from ref. \cite{huang_simulating_2019}. Copyright 2019, American Physical Society. (c) A schematic of a two-photon quantum walk involving only beam splitters. (d) Quantum states are sent through the designated beam splitter to generate a Bell state. Half of it interferes with Alice’s part. The other half of the Bell state is the teleported state, which goes to Bob. Reproduced with permission from ref. \cite{polacchi_teleportation_2024}. Copyright 2024, under CC BY 4.0 license.}}
\label{fig:qt_tech1}
\end{center}

\newpage

% ----------------------------
\newpage
\maketitle

\centering\section*{For Table of Contents Use Only}

\begin{figure}[!hbt]
	\centering
    \includegraphics[width=8.25cm, height=4.45cm,center]{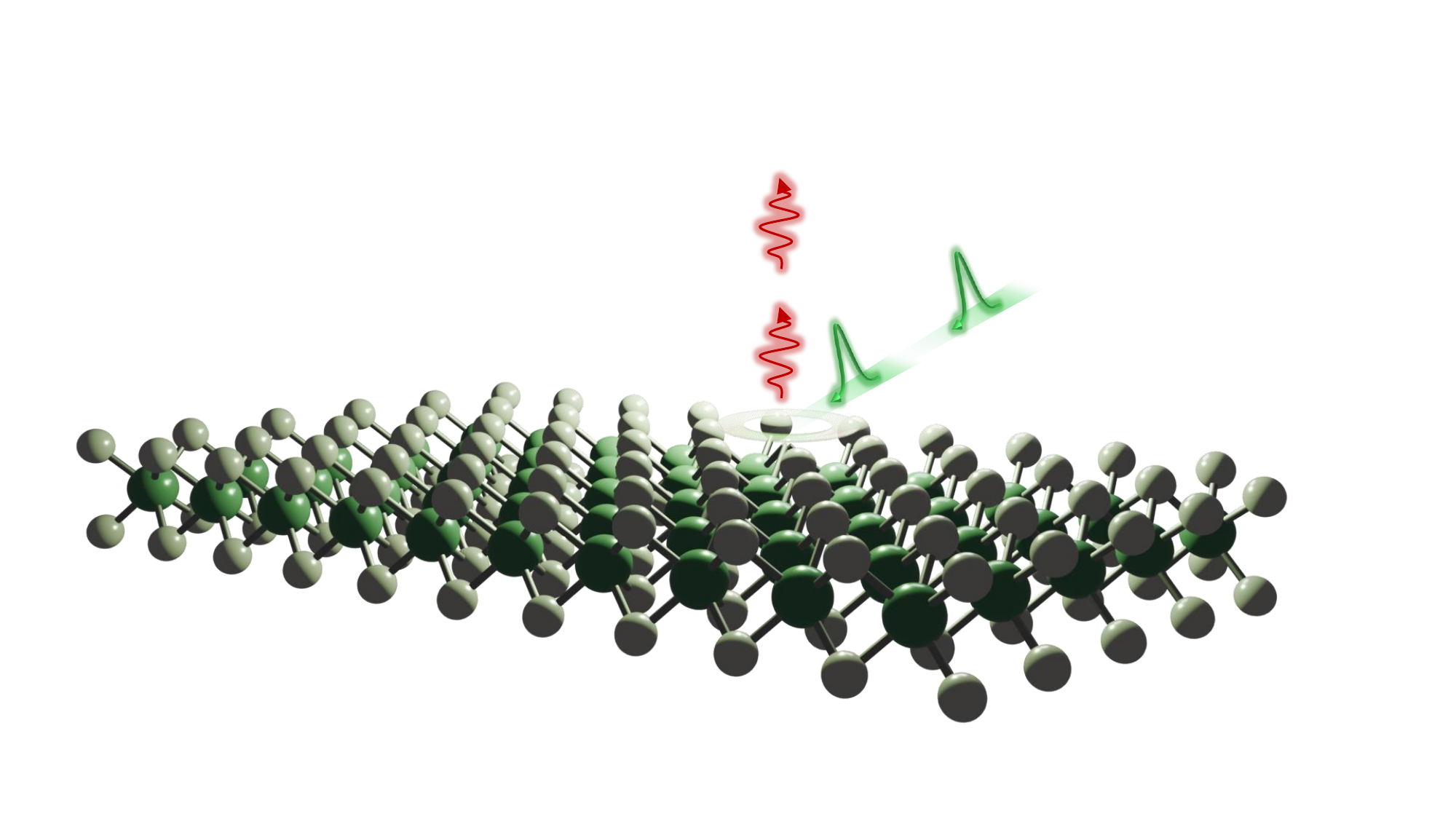}
    \caption*{\revised{This schematic represents a monolayer WSe$_2$ lattice having a selenium (Se) vacancy as a localized defect site. Upon pulsed laser excitation, the defect hosts a quantum emitter that emits single photons. This illustrates defect-induced single photon emission in two-dimensional TMDC materials.}}
    % \label{fig:TOC}
\end{figure}

\phantomsection % Ensures the hyperlink points to the top of this page
\label{page:For_Table_of_Contents_Use_Only}

\end{document}